\documentclass[prb,aps,amsmath,amssymb,showpacs,twocolumn]{revtex4-1}

\usepackage{amssymb}
\usepackage{graphicx}
\usepackage{float}
\usepackage{dcolumn}
\usepackage{bm}
\usepackage{subfigure}
\usepackage{color}

\begin{document}

\title{Adiabatic continuity between Hofstadter and Chern insulator states}

\author{Ying-Hai Wu,$^1$ J. K. Jain,$^1$ and Kai Sun$^{2,3}$}
\affiliation{$^1$ Department of Physics, 104 Davey Lab, The Pennsylvania State University, University Park, PA 16802
}
\affiliation{$^2$
Condensed Matter Theory Center and Joint Quantum Institute,
Department of Physics, University of Maryland, College Park, MD 20742
}
\affiliation{$^3$
Department of Physics, University of Michigan, Ann Arbor, MI 48109
}
\date{\today}

\begin{abstract}
We show that the topologically nontrivial bands of Chern insulators are adiabatic cousins of the Landau bands of Hofstadter lattices. We demonstrate adiabatic connection also between several familiar fractional quantum Hall states on Hofstadter lattices and the fractional Chern insulator states in partially filled Chern bands, which implies that they are in fact different manifestations of the same phase. This adiabatic path provides a way of generating many more fractional Chern insulator states and helps clarify that nonuniformity in the distribution of the Berry curvature is responsible for weakening or altogether destroying fractional topological states.
\end{abstract}
\maketitle

\section{Introduction}

The phenomena of integer and fractional quantum Hall (FQH) effects have motivated remarkable developments. Of particular significance in this context is the topological interpretation of these effects. Thouless, Kohmoto, Nightingale and den Nijs \cite{Thouless} considered electrons in a periodic lattice exposed to a magnetic field, and showed that the Hall conductance of a filled band is related to the first Chern number $C$. Specifically, the Bloch wave function for a magnetic unit cell has the form
\begin{equation}
|\Psi^n(\mathbf{k})\rangle = e^{i\mathbf{k}\cdot\mathbf{r}} |u^n(\mathbf{k})\rangle
\end{equation}
where $\mathbf{k}$ is the wave vector in the first Brillouin zone, and $n$ is the band index. One defines the non-Abelian Berry connection as 
\begin{equation}
\mathbf{\cal A}^{mn}_{\mu}(\mathbf{k})=i\langle u^m(\mathbf{k})| \partial_{\mu} |u^n(\mathbf{k})\rangle
\end{equation}
where $\partial_{\mu}$ is the shorthand notation for $\partial/\partial{\mathbf{k}_{\mu}}$. The Berry curvature $\mathbf{\cal F}$ is then defined as 
\begin{equation}
{\cal F}_{\mu\nu}(\mathbf{k})=\partial_{\mu} \mathbf{\cal A}_{\nu} - \partial_{\nu} \mathbf{\cal A}_{\mu} + i \left[{\cal A}_{\mu}, {\cal A}_{\nu} \right]
\end{equation}
Thouless {\em et al.} showed that the Hall conductance is given, in units of $e^2/h$, by the Berry curvature integrated over the Brillouin zone:
\begin{equation}
\sigma_{\rm H} = \frac{1}{4\pi} \int_{\rm BZ} d^2k \epsilon^{\mu\nu} {\rm Tr} \left[ {\cal F}_{\mu\nu}(\mathbf{k}) \right] = C
\end{equation}
where the trace is over the occupied bands. The Chern number is a topological index provided the sum is over filled bands and the Fermi level lies in a gap. If there is only one occupied band, as we will assume below in this paper, the above expressions simplify to
\begin{eqnarray}
\mathbf{\cal A}_{\mu}(\mathbf{k})=i\langle u(\mathbf{k})| \partial_{\mu} |u(\mathbf{k})\rangle \\
{\cal F}(\mathbf{k})=\partial_{1} \mathbf{\cal A}_{2} - \partial_{2} \mathbf{\cal A}_{1} \\
C = \frac{1}{2\pi} \int_{\rm BZ} d^2k {\cal F}(\mathbf{k})
\end{eqnarray}
Ref.~[\onlinecite{Thouless}] thus gave a topological interpretation of the quantized Hall conductance and also clarified that the essential property of a Landau level that distinguishes it from an ordinary band is its nonzero Chern number. Subsequently, Haldane \cite{Haldane1} showed that a uniform magnetic field is not required to produce bands with nonzero Chern numbers. For this purpose he constructed an explicit model of an electron hopping on the honeycomb (graphene) lattice, with complex hopping matrix elements; this model has no {\em net} magnetic field (although it has staggered magnetic field) but produces bands with nonzero Chern numbers in certain regions of the parameter space. This system has integrally quantized Hall conductance in the absence of a uniform magnetic field. Systems with bands of nontrivial topology (nonzero Chern number) in the absence of a uniform magnetic field are now called ``Chern insulators," to distinguish them from the Landau levels that occur in the presence of a uniform magnetic field. A number of other models have been proposed for Chern insulators, some of which have nearly flat bands \cite{Tang,Sun,Neupert}. 

The next natural question is whether Chern bands can also support FQH-like states; that is incompressible states in a partially filled Chern band with a fractionally quantized Hall conductance. Such states will obviously require interactions, and have been dubbed fractional Chern insulator (FCI) states. Exact numerical diagonalizations have demonstrated FCI states at filling factors $\nu=1/3$,~\cite{Sheng, Regnault, Wu1} $1/2$,~\cite{Bernevig} $2/5$ and $3/7$~\cite{Liu} for fermions, and $1/2$,~\cite{Wang1} $1$,~\cite{Wang2} and $2/3$ \cite{Liu} for bosons. These states require specific forms of interaction, and sometimes fine tuning of parameters. Trial wave functions for FCIs have been proposed \cite{Qi,Lu,Wu2}. Flat-band models with $C>1$ have also been constructed and produce strongly-interacting topological states \cite{Ran,Wang,Trescher,Bergholtz,Yang}. 

While the FCI states appear similar to the FQH states found in the lowest Landau level, no direct connection between them has yet been established. It remains unclear why some fractions occur while others do not, and what is the role of lattice symmetry and the type and range of the interaction in establishing various FCI states. It also remains unclear to what extent the extensive physics of the FQH effect and composite fermions is possible in Chern insulators. Progress in this direction has been made by Murthy and Shankar \cite{Murthy}, who exploit the modified algebra of the density operator projected into the lowest Chern band to motivate composite fermion physics. 

We address below this issue by demonstrating an adiabatic continuity between the ordinary quantum Hall states in a Landau level and the corresponding states in a Chern insulator. Because the latter are defined on a lattice, we work with a lattice model of electrons in a uniform magnetic field. This problem of Bloch electrons in a magnetic field was studied in a number of papers, including those by Peierls,~\cite{Peierls} Harper,~\cite{Harper} Wannier,~\cite{Wannier}, Azbel,~\cite{Azbel}, and Hofstadter,~\cite{Hofst}, with the last article presenting the band structure in a pictorially appealing form that is now known as the Hofstadter butterfly. For appropriately chosen flux per plaquette, the low-lying Bloch bands of this system are essentially Landau levels; they approximate Landau levels of the continuum very accurately for a sufficiently fine lattice. We call them ``Hofstadter bands," and the filled band states ``Hofstadter insulators." For a given Hofstadter lattice, not all FQH states of the continuum will occur, and which ones survive is an interesting problem in its own right, but will not be addressed in this article (some work along these lines can be found in the literature \cite{Sorensen,Moller2}). However, we can certainly construct a Hofstadter lattice that approximates the continuum arbitrarily closely, by taking the flux per plaquette to be sufficiently small, and thus it produces all of the quantum Hall states seen in continuum. (Strictly speaking, the electrons in GaAs quantum wells are not in a continuum but feel the periodic potential of the lattice.)  We will study a possible adiabatic connection between the quantum Hall states on a Hofstadter lattice and the corresponding states in a Chern insulator.

An intuitive understanding for why an adiabatic connection between a Chern and a Hofstadter lattice may exist can be gained by noting that a Hofstadter insulator in a {\em uniform} magnetic field can be transformed into a Chern insulator in zero {\em net} magnetic field by a simple gauge transformation. For a Hofstadter lattice, the total magnetic field passing through each magnetic unit cell is $2q\pi$ ($q$ is an integer). Here and below, one flux quantum is defined as $\phi_0=2\pi{\hbar}c/e=2\pi$ in units with $\hbar=c=e=1$. Let us now insert a $-2q\pi$ flux at an arbitrary point in each magnetic unit cell to produce a new problem, called Hofstadter$^\prime$ (``Hofstadter prime").  The insertion of the $-2q\pi$ flux in a tight-binding model, however, is simply a gauge choice that has no physical consequence, and hence leaves all properties of the system unchanged: the energy bands of the Hofstadter' lattice are identical to those of the original Hofstadter lattice, and the eigenfunctions of the two are related by a gauge transformation. In particular, the bands of the new lattice continue to have nonzero Chern numbers. At the same time, if we treat the (enlarged) magnetic unit cell as our unit cell, then the total magnetic field through it is zero. The Hofstadter' lattice is thus a Chern insulator. (In fact, this Chern insulator has flat bands and uniform Berry curvature.) Every Hofstadter insulator thus has a corresponding Chern insulator with identical properties. This implies that all of the physics of FQH effect and composite fermions is, in principle, possible for Chern insulators, provided that we allow Chern insulators with a sufficiently complex unit cell. 

For simple Chern lattices, not all FQH states occur. In what follows, we consider certain previously introduced Chern insulator models, construct for each a Hofstadter' lattice whose magnetic unit cell coincides with the unit cells of the Chern insulator, and show, using exact numerical methods, that the familiar FQH states at filling factors $1/3$ and $1/2$ of the Hofstadter' model adiabatically evolve into the corresponding FCI states in the presence of appropriate repulsive interactions. (Because Hofstadter' lattice is trivially related to the Hofstadter lattice, we will dispense with the prime below.)  We show that not only does the ground state evolve in this manner, but so do the quasiholes and the entanglement spectra, lending further credence to such an adiabatic relationship. This demonstrates that the origin of these states is governed by the same underlying physics. Furthermore, this adiabatic connection also enables us to investigate the role of the Berry curvature distribution in the momentum space. We find that nonuniformity in the distribution of the Berry curvature weakens, and can even destroy, FQH states. Our results show that such nonuniformities effectively translate into an enhancement of the residual interaction between composite fermions, and as a result can eliminate states of the sequence $p/(2p\pm1)$ with relatively small gaps (all these fractions would occur for noninteracting composite fermions). Nonetheless, several FQH states are surprisingly robust to nonuniformities of the Berry curvature.

The paper is organized as follows. We present two single-particle tight-binding models with topologically nontrivial lowest bands in Sec. II. The FQH states on these lattice models, with appropriately chosen interactions, are studied in Sec. III. Sec. IV concludes with a discussion of the implications of our results. Since the posting of the first version of this work as arXiv:1207.4439v1, some new results~\cite{Lauchli,Roy,LiuBerg} have appeared, which are also discussed in Sec. IV. 

\section{Lattice Models and Integer Quantum Hall Effect}

We consider two popular models for Chern insulators: the checkerboard and the kagome lattices. In either case, our goal is to write a more general model that extrapolates between a Hofstadter lattice and a Chern insulator lattice. For this purpose we add many more lattice sites to the Chern insulator lattice to create a Hofstadter lattice, and arrange the flux per plaquette so that the Hofstadter lattice has the same magnetic unit cell as the Chern insulator being considered, and also has a net zero magnetic field passing through the magnetic unit cell. With this arrangement both the lattices have the same symmetries (although they have different numbers of bands, because they have different numbers of lattice sites in a unit cell) and it is sensible to envision an adiabatic evolution from one to the other. We first study the single-particle band structures to demonstrate an adiabatic connection between a Landau level and a Chern band for the two lattice models mentioned above. 

\begin{figure}
\centering
\subfigure[]{\includegraphics[width=0.23\textwidth]{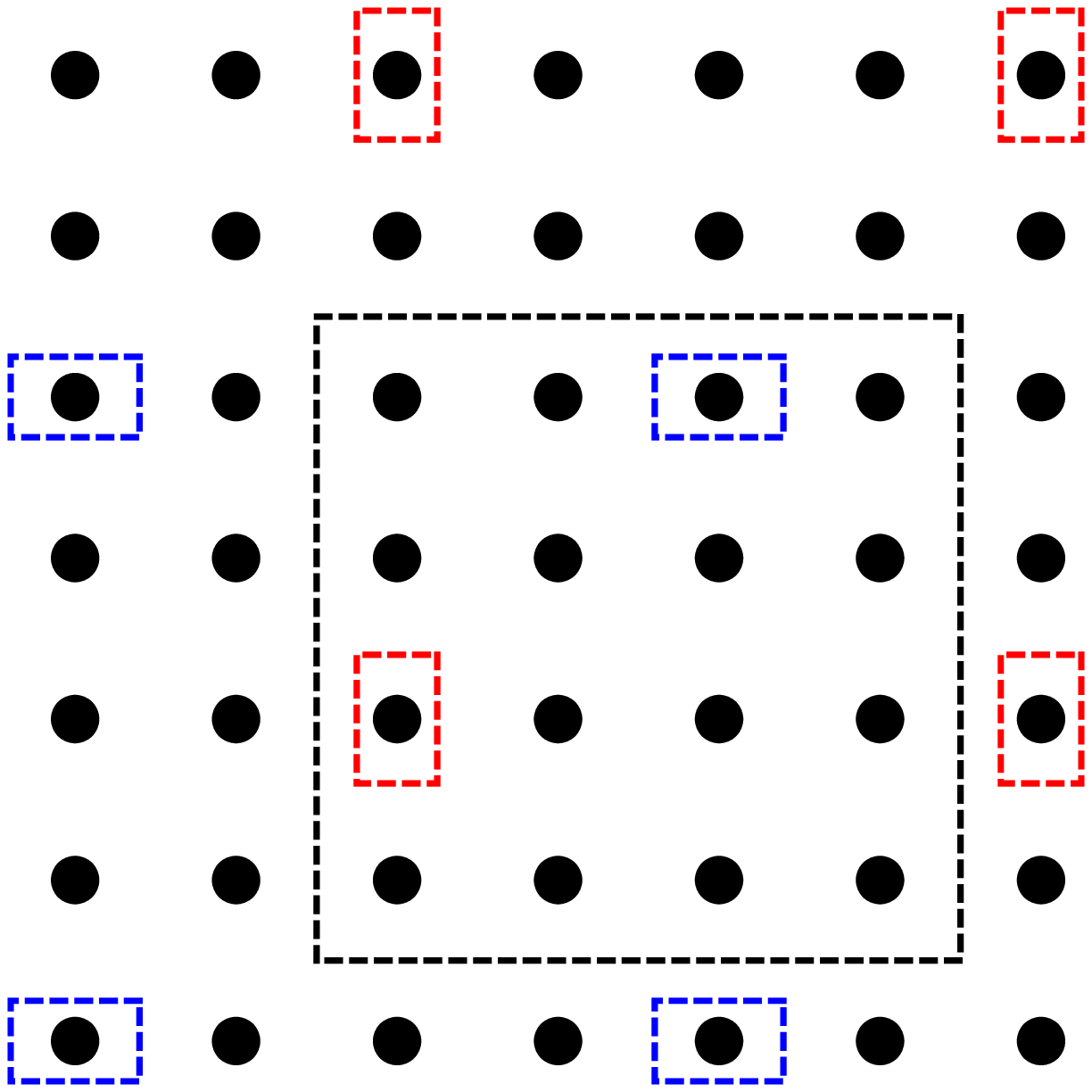}}            
\subfigure[]{\includegraphics[width=0.23\textwidth]{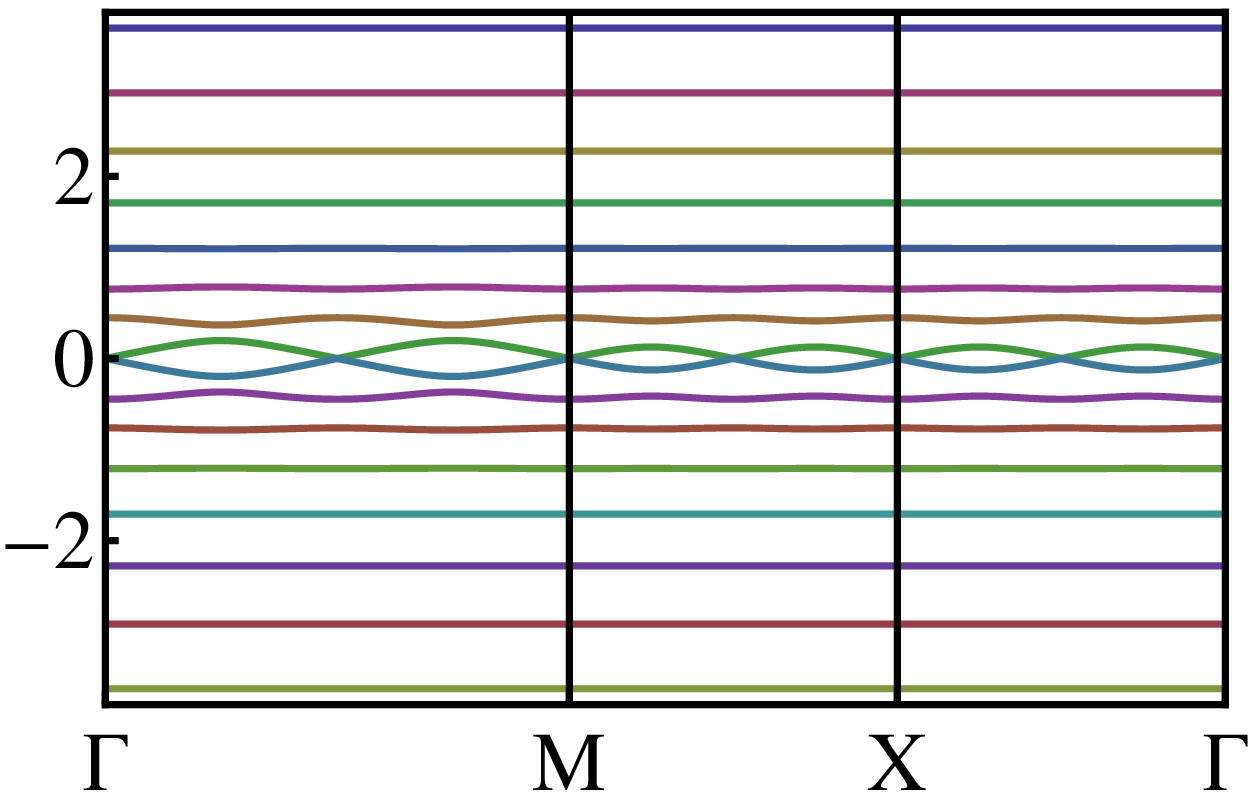}}
\centering
\subfigure[]{\includegraphics[width=0.23\textwidth]{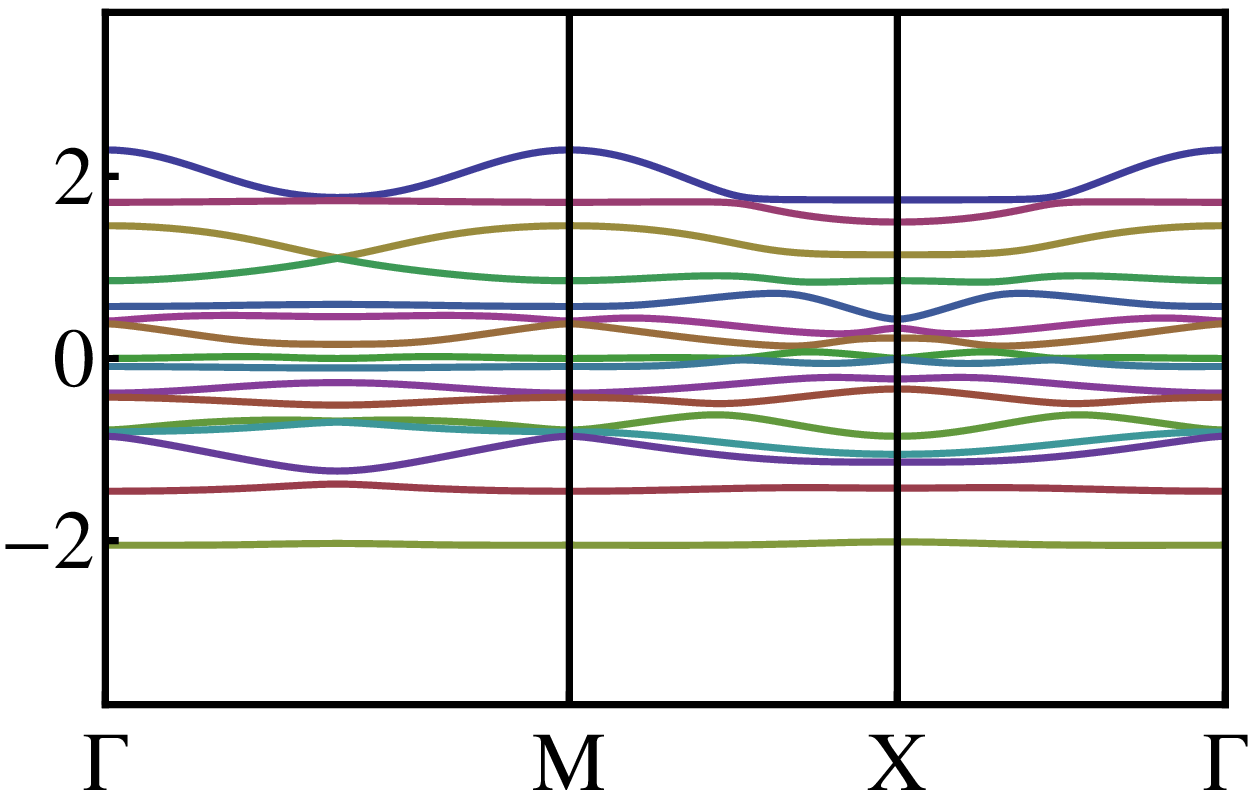}}            
\subfigure[]{\includegraphics[width=0.23\textwidth]{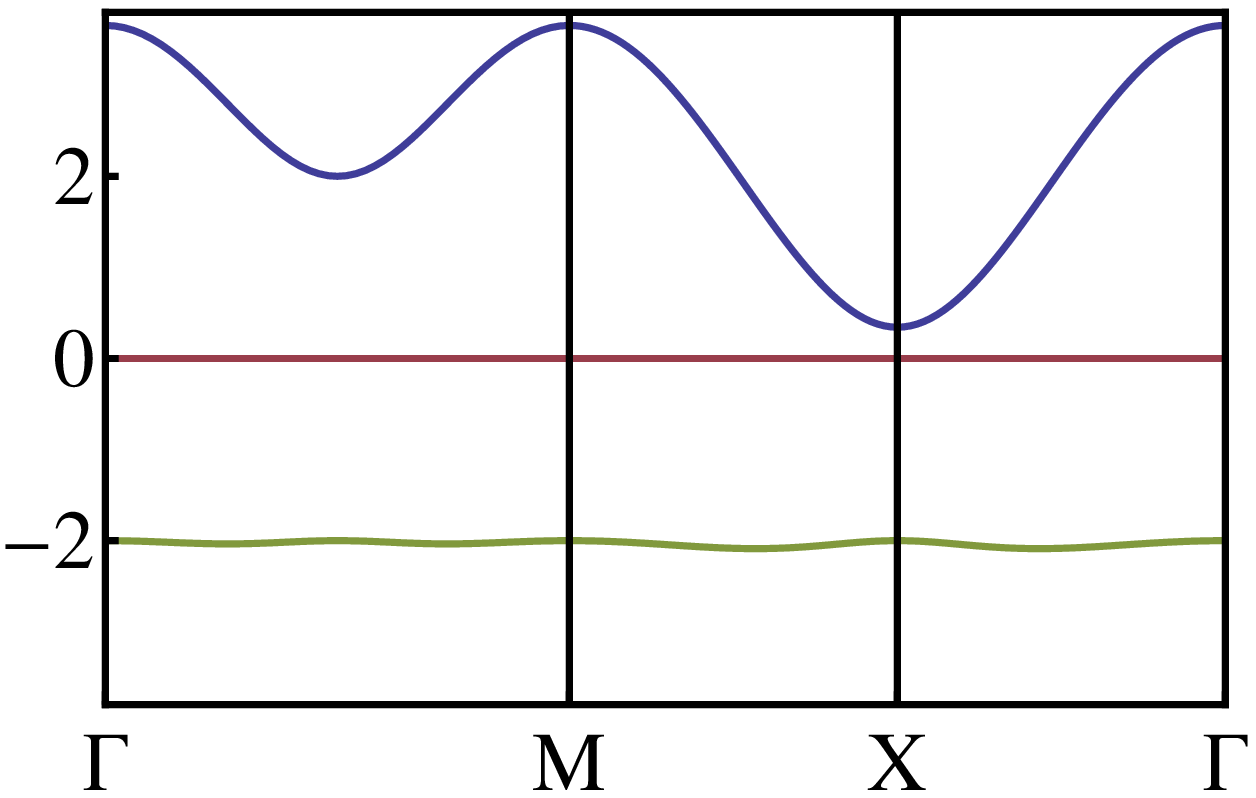}}
\caption{Lattice model and band structure for the square-checkerboard lattice model. Panel (a) shows the lattice structure. The black dots shows the lattice sites in a Hofstadter lattice with flux $2\pi/16$ through each plaquette. The dashed black square marks a magnetic unit cell containing $16$ sites. The sites marked by the dashed rectangles form a checkerboard lattice where the two different orientations of the rectangles represent two sublattices. The Hamiltonian $H_{\rm sq-cb}(R)$ in Eq.~(\ref{squarechecker}) interpolates between the Hofstadter and Chern insulator limits as $R$ is varied from 0 to 1. Panels (b-d) show the band structures at three values of $R$ ($0$, $0.5$ and $1$, respectively) along the contour $\Gamma \rightarrow M \rightarrow X\rightarrow \Gamma $ in the momentum space. In panel (b), flat Landau levels carry Chern number $C=1$ while the two non-flat bands at the middle have a total $C=-14$. In panel (d), the top and bottom bands of the checkerboard model have nontrivial Chern numbers $C=\mp1$.}
\label{LatticeBandChecker}
\end{figure}

\begin{figure}
\includegraphics[width=0.40\textwidth]{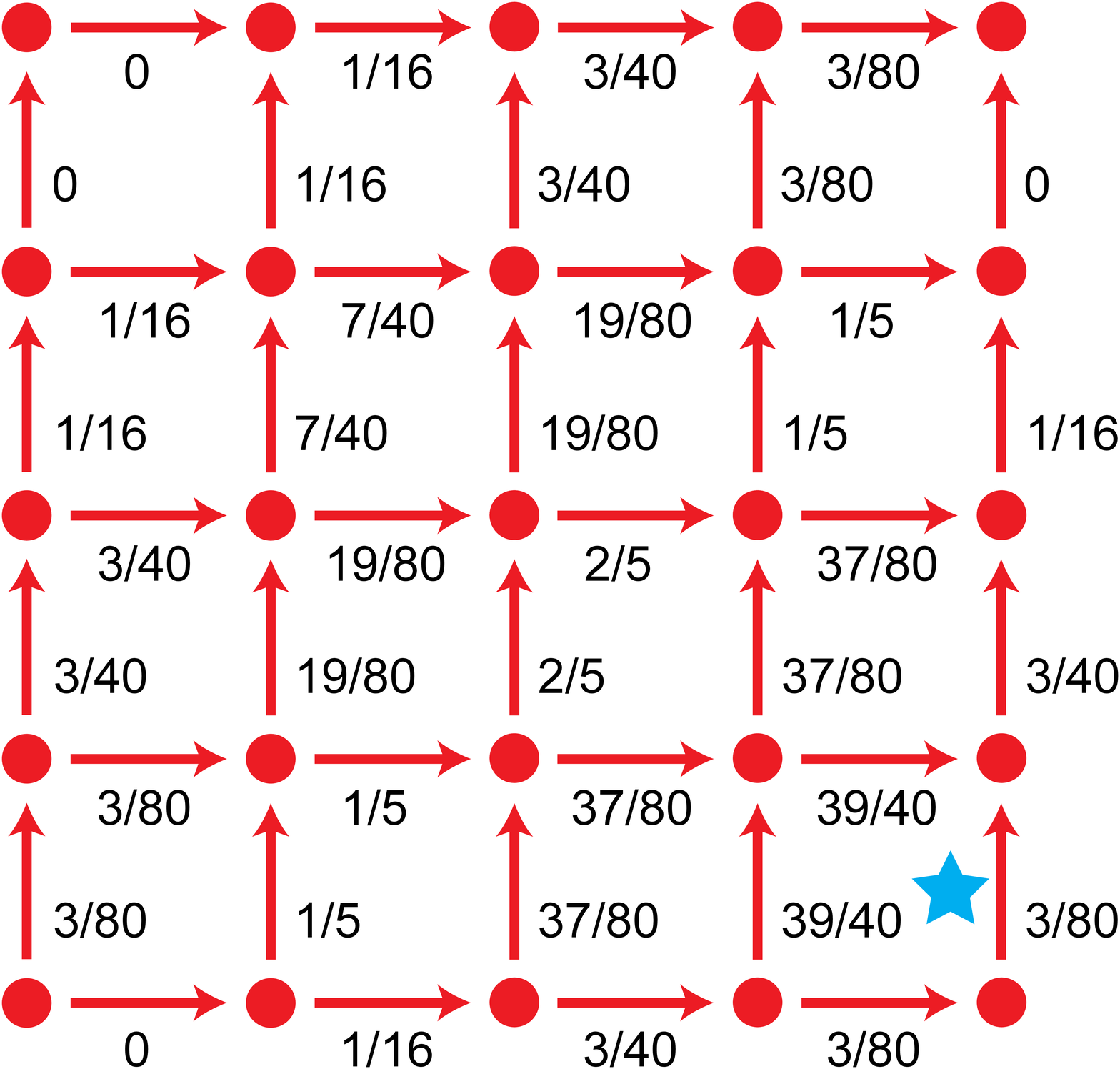}
\caption{Phases of the square lattice Hofstadter model. The numbers and arrows indicate the hopping phases along the bonds in units of $\pi$, and the star marks the plaquette where a $-2\pi$ flux is inserted.}
\label{SquareLatticePhase}
\end{figure}

\begin{figure}
\subfigure[]{\includegraphics[width=0.32\textwidth]{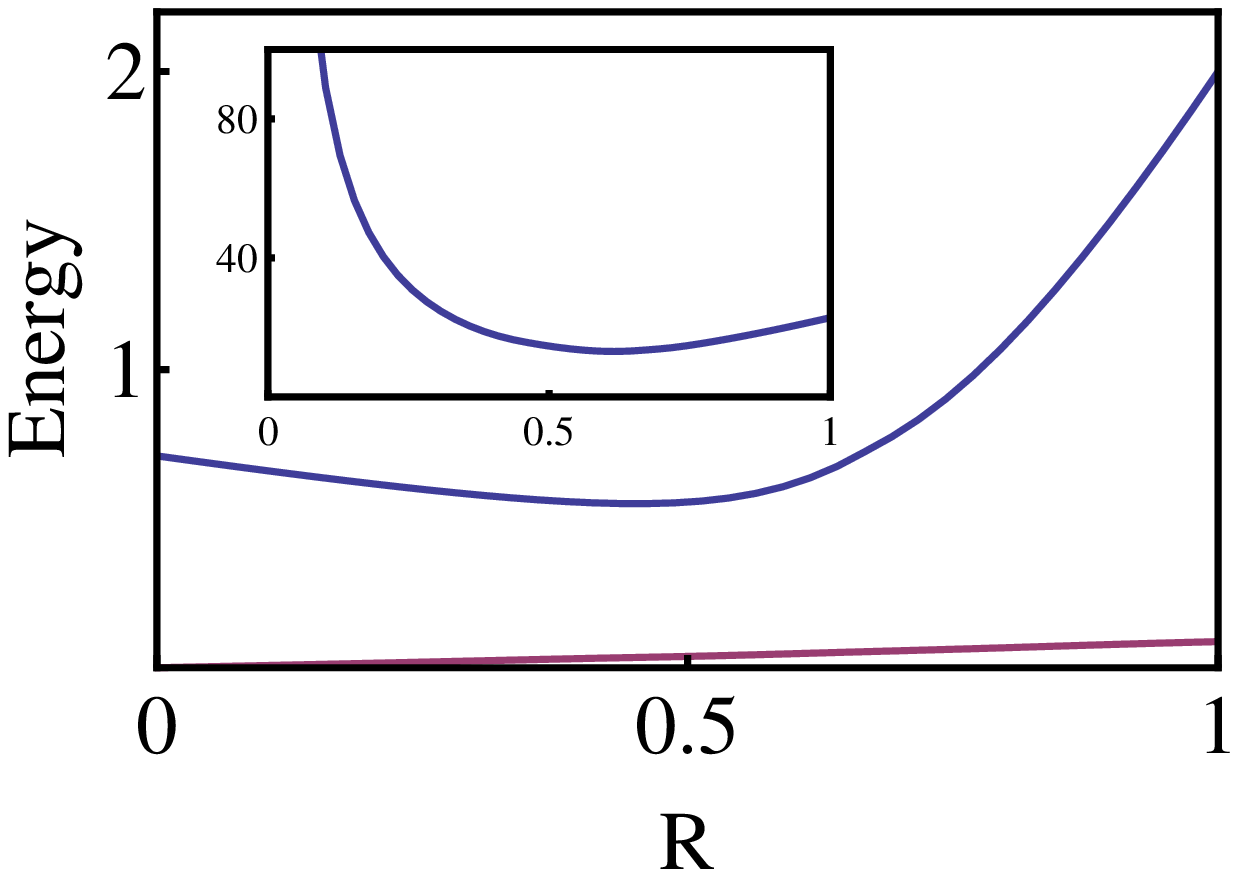}}
\subfigure[]{\includegraphics[width=0.32\textwidth]{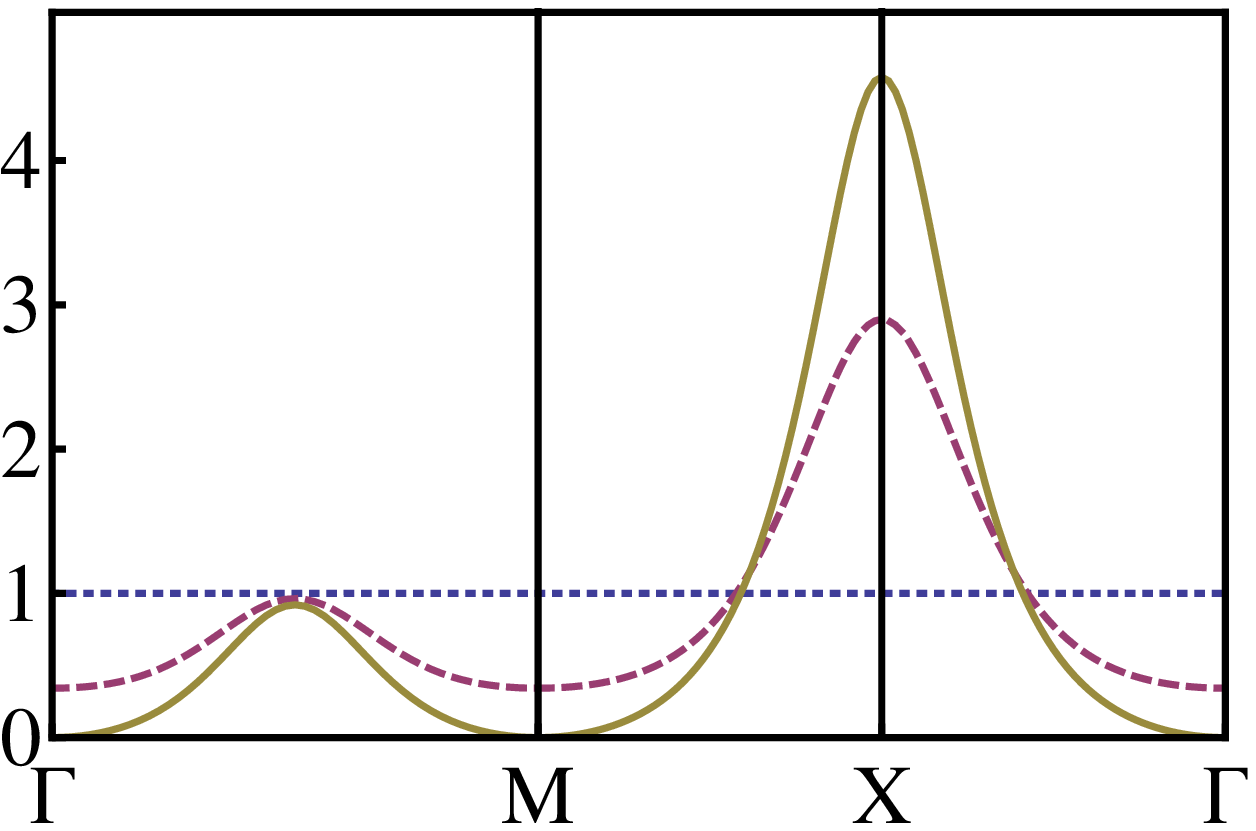}}
\caption{Square-checkerboard hybrid lattice (a) The single-particle band gap (top curve) and bandwidth (lower curve) as a function of $R$. The flatness ratio (band gap over bandwidth) is shown as the inset. (b) The Berry curvature at different {\bf k}-points with $R=0.0$, $0.5$ and $1.0$  (dotted, dashed, and solid lines, respectively).}
\label{FlatBerryChecker}
\end{figure}

\subsection{Square-checkerboard Hybrid Lattice}

The checkerboard lattice model was considered in Refs.~\onlinecite{Sun} and~\onlinecite{Sheng}. It supports a topologically nontrivial band in the presence of nearest-neighbor, next-nearest-neighbor, and next-next-nearest-neighbor hopping terms, with the nearest neighbor hopping carrying a nonzero phase. By appropriate choice of parameters, the lowest band can be made very nearly flat, which is an important consideration for the discussion of the FCI states, which require that the interaction energy dominate the kinetic energy. The checkerboard lattice is shown in Fig.~\ref{LatticeBandChecker}(a) by the encircled dots, with its two sublattices marked by blue and red rectangles. The checkerboard Hamiltonian $H_{\rm cb}$ is given by
\begin{eqnarray}
H_{\rm cb} &=& - t \sum_{\left<ij\right>} e^{i\phi_{ij}} c^\dagger_i c_j - t_1 \sum_{\left<\left<ij\right>\right>} {\rm sgn} \;\; c^\dagger_i c_j \nonumber \\
&-& t_2 \sum_{\left<\left<\left<ij\right>\right>\right>} c^\dagger_i c_j + {\rm h.c.}
\end{eqnarray}
where $\phi_{ij}$ is the phase acquired during hopping between nearest neighbors $\left<ij\right>$, the $\rm sgn$ is positive (negative) for next-nearest-neighbor hoppings between sites encircled by blue (red) rectangles, and $\left<\left<\left<ij\right>\right>\right>$ denotes next-next-nearest neighbors.

As shown in Fig.~\ref{LatticeBandChecker}(a), we embed the checkerboard lattice inside a square Hofstadter lattice (all black dots) with only nearest neighbor hopping. It is assumed that each square of this lattice has a magnetic flux $2\pi/n$ passing through it ($n=16$ for Fig.~\ref{LatticeBandChecker}(a)), with the exception of one plaquette (indicated with a star in Fig.~\ref{SquareLatticePhase}) which has an additional $2\pi$ flux passing through it in the opposite direction (so the net flux is $-(15/16)2\pi$ through this plaquette), so as to make the total flux through the magnetic unit cell equal to zero. (In other words, the lattice is what we had called Hofstadter$^\prime$.) The hopping matrix elements are complex, with our choice of phases shown in Fig.~\ref{SquareLatticePhase}. The phases are chosen to obey periodic boundary conditions, but that does not fix them uniquely; we further impose the convention that the phase coming up to a site is the same as the phase going out of it toward right, as shown in the figure. It is straightforward to verify that the phases correspond to a flux of $2\pi/16$ through each square, except for the starred one which has an additional flux of $-2\pi$ through it. The Hamiltonian of this Hofstadter model on square lattice is denoted by $H_{\rm sq}$. 

We note that we could have taken a finer lattice with a larger number ($n$) of squares per unit cell; Landau levels are recovered in the weak lattice limit of $n\gg 1$. For our purposes the current choice with 16 squares in a magnetic unit cell will suffice, as seen below in the explicit numerical calculations. It is also noteworthy that if we only had the Hofstadter lattice (no checkerboard lattice), then we could have chosen a linear magnetic unit cell with 16 sites along a single line; this magnetic unit cell arises naturally in the Landau gauge, and was the choice made by Hofstadter. However, our objective of matching the Hofstadter problem to the checkerboard problem forces us to choose identical unit cells for both of them. We are not able to explicitly write a real space gauge potential that would produce the desired phases for the hopping matrix elements, but an appropriate gauge choice is made directly at the level of the phases of the hopping matrix elements. 

Having defined the checkerboard and the Hofstadter Hamiltonians separately, we now define an interpolating Hamiltonian
\begin{equation}
H_{\rm sq-cb}(R)=(1-R) H_{\rm sq} + R H_{\rm cb},
\label{squarechecker}
\end{equation}
which evolves continuously from Hofstadter to checkerboard as $R$ increases from $0$ to $1$. This interpolation scheme selects one specific path in the parameter space connecting the Hofstadter and the checkerboard Hamiltonians. We shall see that this path will suffice for demonstrating adiabatic continuity for many situations. By Fourier transformation, the above Hamiltonian can be converted into its momentum-space form $H_{\rm sq-cb}=\sum_{\mathbf{k}}\sum_{\alpha\beta} c^\dagger_{\mathbf{k}\alpha} {\cal H}^{\alpha\beta}_{\rm sq-cb}(\mathbf{k}) c_{\mathbf{k}\beta}$ and the nonzero components of ${\cal H}^{\alpha\beta}_{\rm sq-cb}$ are given in the Appendix, where $\alpha,\beta=0,1,\cdots,15$. Diagonalization of this 16 $\times$ 16 Hamiltonian produces the eigenstates and eigenvalues for the 16 bands as a function of the two-dimensional wave vector $\mathbf{k}$ over the entire Brillouin zone. 

We have considered a large number of values of $R$ to reach our conclusions noted below, but, for brevity, we will show results only for $R=0$, $R=1$ and an intermediate value $R=0.5$. Fig.~\ref{LatticeBandChecker} (b), (c) and (d) show the band structure at these $R$ values. Explicit calculation shows that the lowest band always remains gapped as a function of $R$, which implies that the lowest ``Landau level" of the Hofstadter model adiabatically evolves into the lowest band of the Chern insulator, carrying along its Chern number. In Fig.~\ref{FlatBerryChecker} (a), we show the band gap and the band width of as functions of $R$. The Chern insulator band can thus be considered a renormalized Landau level. The integer quantum Hall states in the Chern insulator thus are adiabatically connected to their counterparts in Landau level systems. Even though the flatness ratio (band gap divided by band width) remains large, the Berry curvature changes drastically as shown in Fig.~\ref{FlatBerryChecker} (b). 

One may note that as $R$ approaches 1, 14 of the 16 bands of the Hofstadter lattice become degenerate at zero energy, reflecting the fact that 14 of the lattice sites in each unit cell essentially drop out of the problem, being completely disconnected from other sites. The band structure at $R=1$ thus contains two dispersive bands and 14 degenerate bands at $E=0$. This drastic rearrangement of higher bands underscores the nontriviality of the adiabatic evolution of the lowest band.

\begin{figure}
\centering
\subfigure[]{\includegraphics[width=0.23\textwidth]{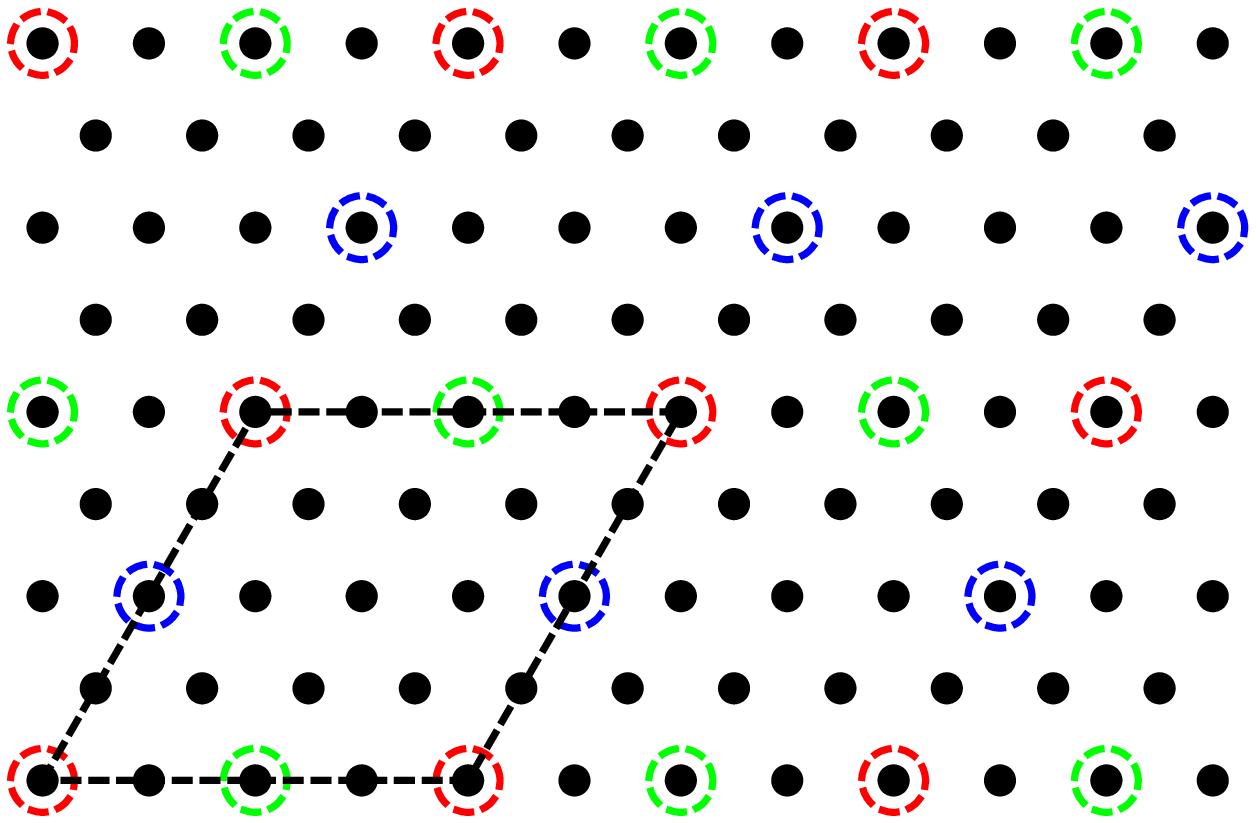}}
\subfigure[]{\includegraphics[width=0.23\textwidth]{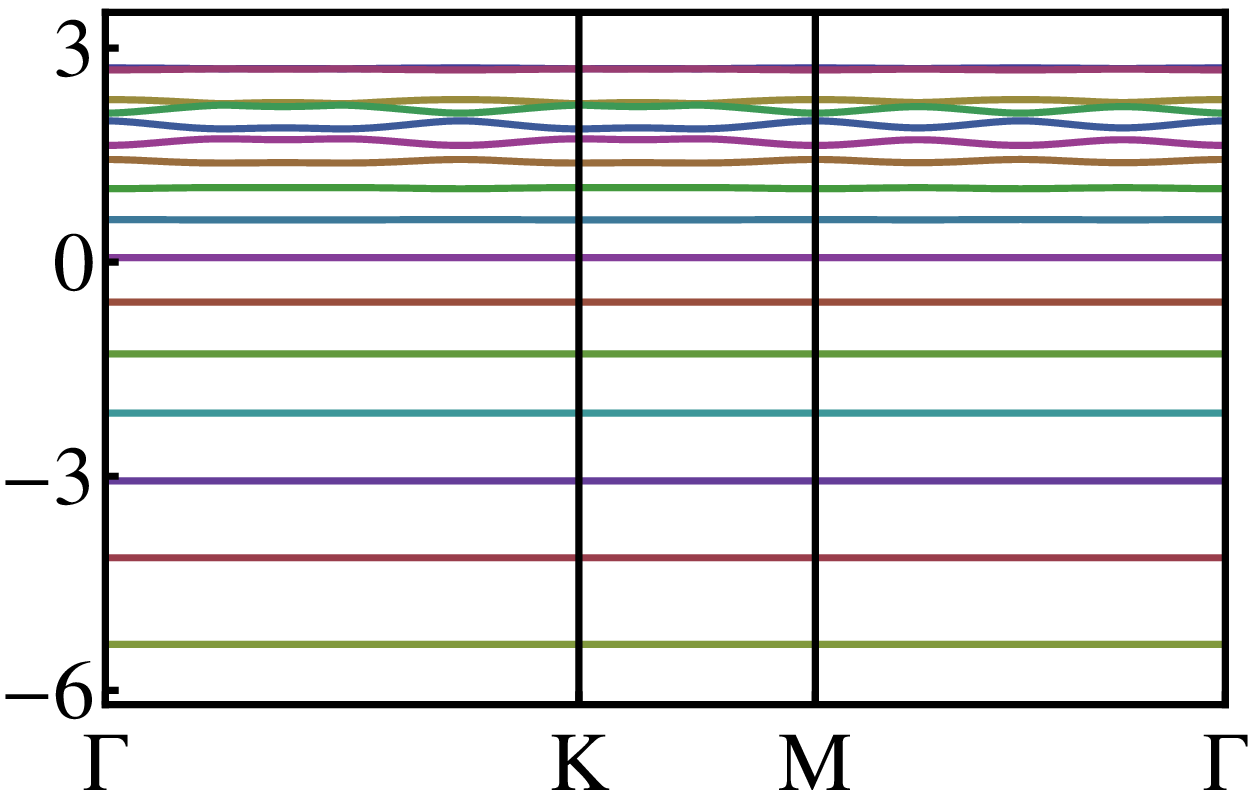}}
\centering
\subfigure[]{\includegraphics[width=0.23\textwidth]{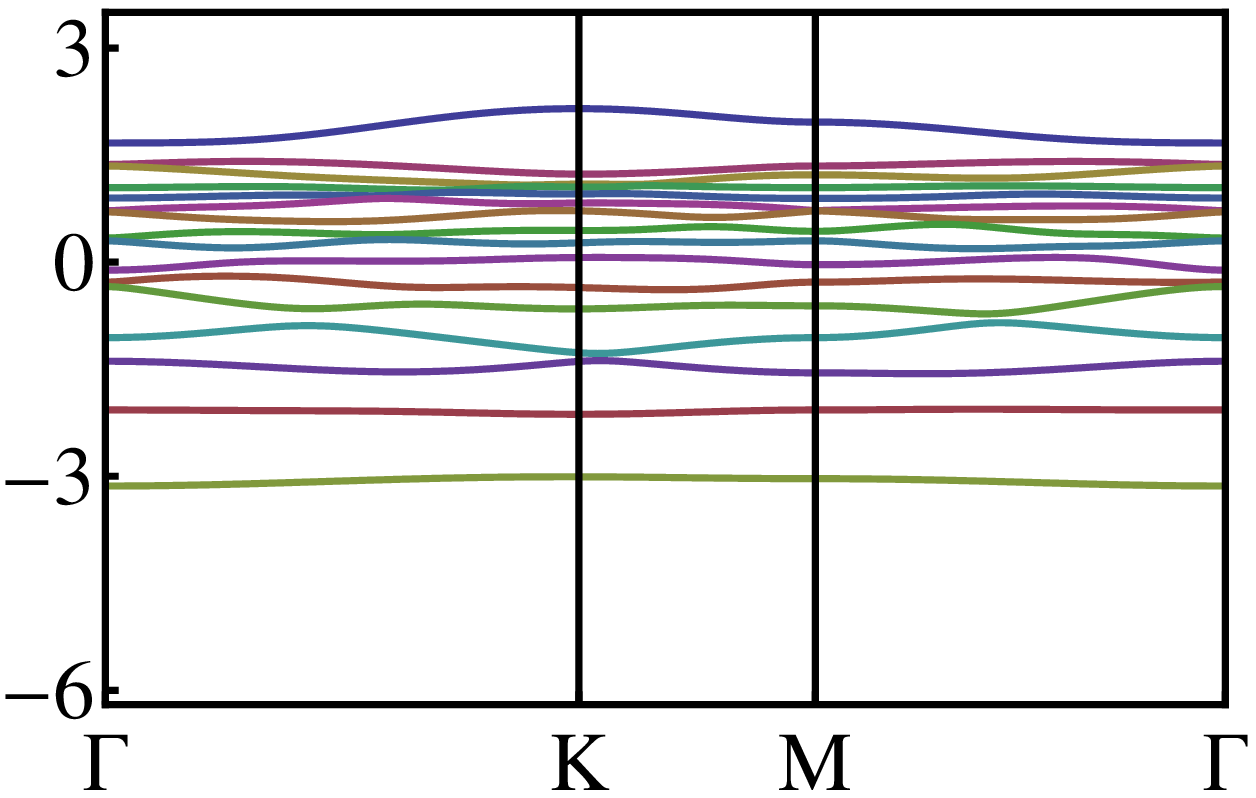}}
\subfigure[]{\includegraphics[width=0.23\textwidth]{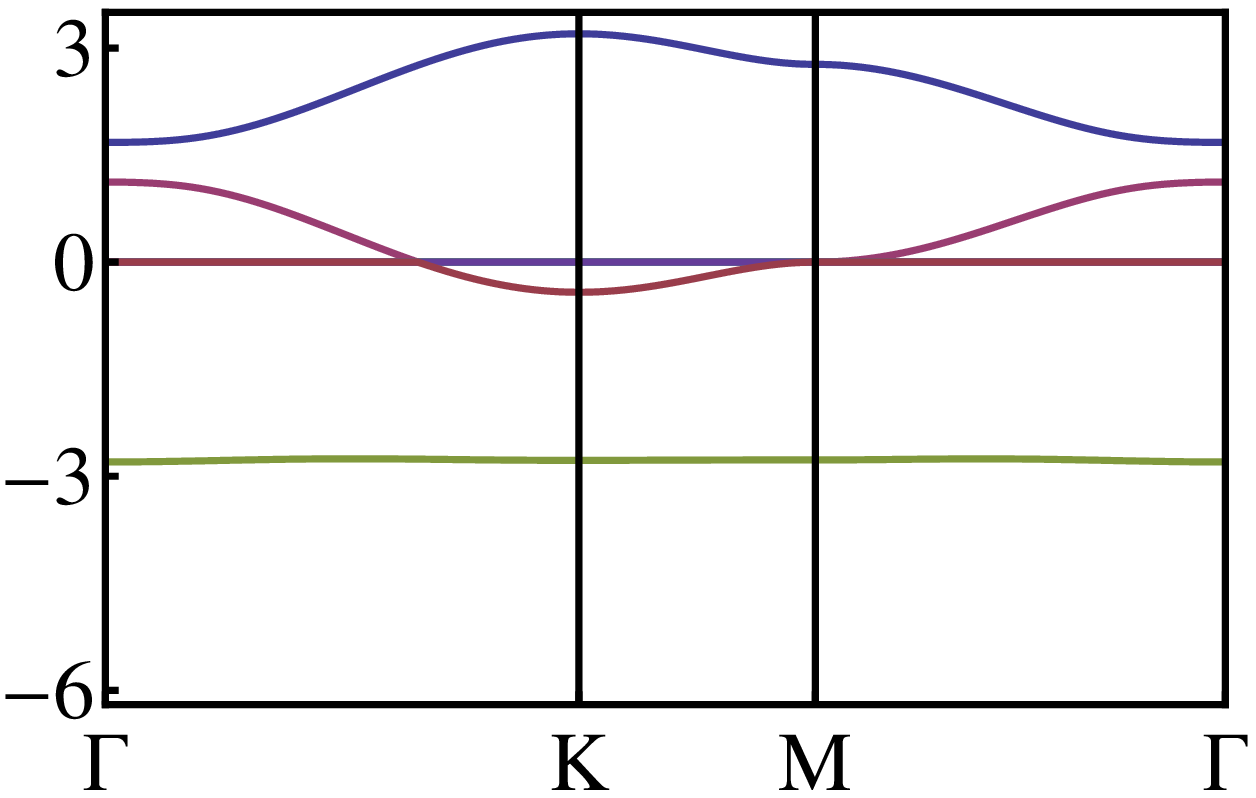}}
\caption{Lattice model and band structure for the triangnular-kagome  lattice model. Panel (a) shows the lattice structure. The black dots show the lattice sites in a triangular Hofstadter lattice with flux $\pi/16$ in each triangle. The dashed lines mark a magnetic unit cell, which contains $16$ sites. The sites marked by the dashed circles form a kagome  lattice where the three different colors represent the three sublattices. Panels (b-d) shows the band structures at different values of $R$ ($0.0$, $0.5$, and $1$, respectively) along the contour $\Gamma \rightarrow K \rightarrow M \rightarrow \Gamma$ in momentum space.}
\label{LatticeBandKagome}
\end{figure}

\begin{figure}
\includegraphics[width=0.47\textwidth]{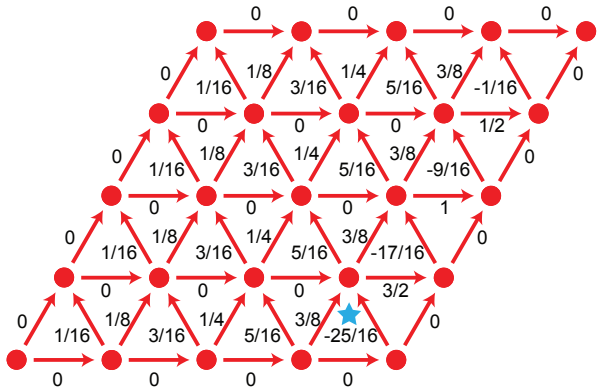}
\caption{Phases of triangular lattice Hofstadter model. With the exception of the outermost hopping bonds, the phase factor associated with a bond, in units of $\pi$, is indicated by the arrow on it and the number either below or to the left of it, and the star marks the triangle where a $-2\pi$ flux is inserted.}
\label{TriangleLatticePhase}
\end{figure}

\begin{figure}
\subfigure[]{\includegraphics[width=0.32\textwidth]{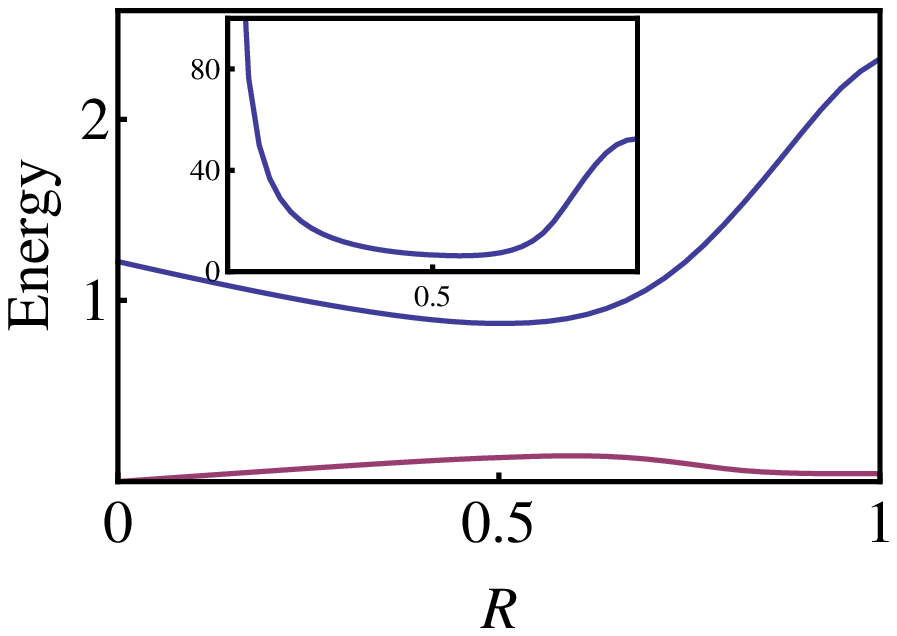}}
\subfigure[]{\includegraphics[width=0.32\textwidth]{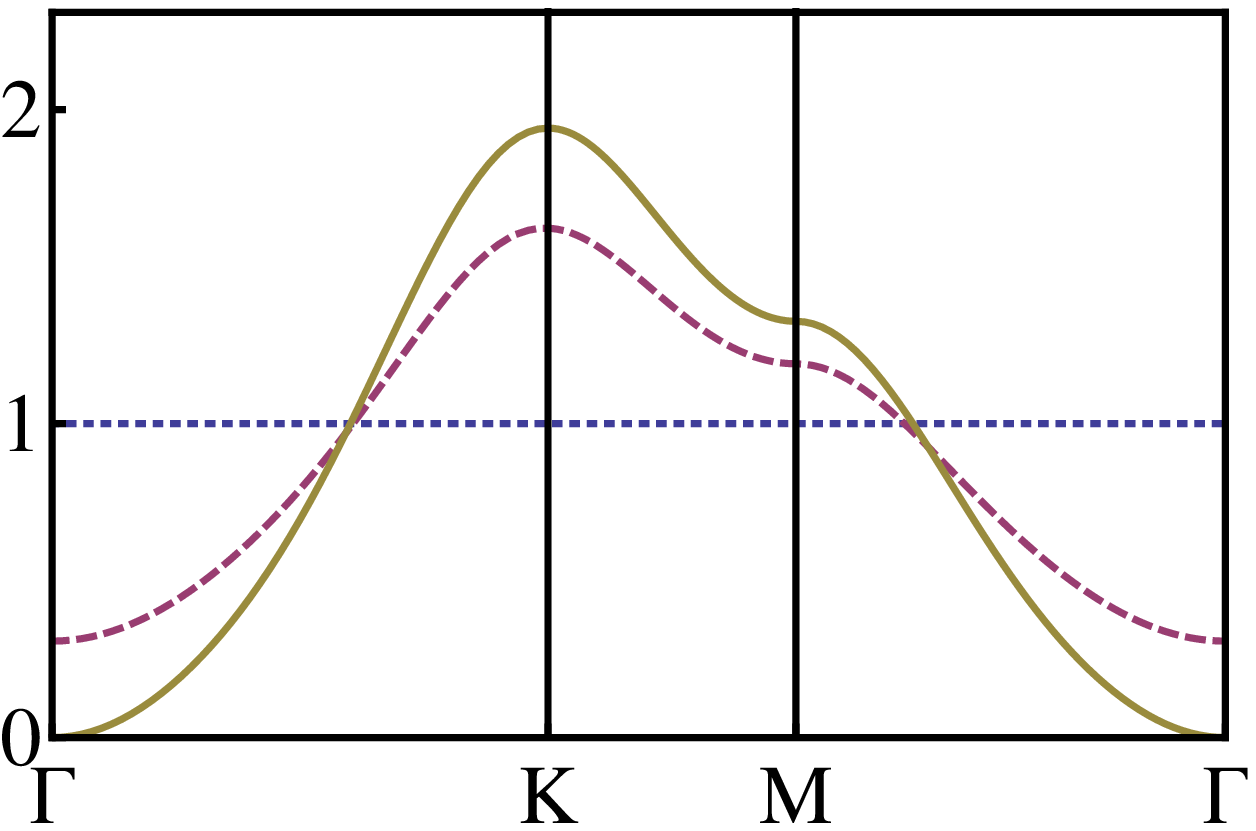}}
\caption{Triangular-kagome hybrid lattice (a) The single-particle band gap (top curve) and bandwidth (lower curve) as a function of $R$. The flatness ratio (band gap over bandwidth) is shown in the inset. (b) The Berry curvature at different {\bf k}-points with $R=0.0$, $0.5$ and $1.0$ (dotted, dashed and solid lines respectively).}
\label{FlatBerryKagome}
\end{figure}

\subsection{Triangular-kagome Hybrid Lattice}

The kagome lattice was introduced in Ref.~[\onlinecite{Tang}], and has been found to be a excellent platform of FCI states \cite{Wu1,Liu}. In Fig.~\ref{LatticeBandKagome}, the kagome lattice is indicated with the encircled dots, with its three sublattices marked by blue, red and green circles. In the original proposal~\cite{Tang}, with complex hopping terms, nearly flat lowest band with Chern number $1$ can be obtained by tuning parameters. The Hamiltonian of this model is 
\begin{eqnarray}
H_{\rm ka} = - t_1 \sum_{\left<ij\right>} c^\dagger_i c_j - t_2 \sum_{\left<\left<ij\right>\right>} c^\dagger_i c_j + {\rm h.c.}
\end{eqnarray}
where $\left<ij\right>$ denotes nearest neighbors and $\left<\left<ij\right>\right>$ next-nearest neighbors and $t_1$ and $t_2$ are complex hopping coefficients. We also embed the kagome lattice inside a Hofstadter lattice, which is chosen to be a triangular lattice with sixteen lattice sites in each magnetic unit cell, as shown in Fig.~\ref{LatticeBandKagome}. There is $\pi/16$ magnetic flux passing through each triangle as shown in Fig.~\ref{TriangleLatticePhase} except the one indicated with a star, where an additional $2\pi$ flux passes through it in the opposite direction. As in the square-checkerboard lattice case, we define the gauge through an explicit choice of the phases as shown in Fig.~\ref{TriangleLatticePhase}. The triangular Hofstadter lattice Hamiltonian $H_{\rm tri}$ with nearest-neighbor hopping carrying these phases give almost flat lowest band and nearly constant Berry curvature. An interpolating Hamiltonian between the triangular and kagome limits is defined as
\begin{equation}
H_{\rm tri-ka}(R)=(1-R) H_{\rm tri} + R H_{\rm ka}
\label{trianglekagome}
\end{equation}
The momentum-space Hamiltonian is given by $H_{\rm tri-ka}=\sum_{\mathbf{k}}\sum_{\alpha\beta} c^\dagger_{\mathbf{k}\alpha} {\cal H}^{\alpha\beta}_{\rm tri-ka}(\mathbf{k}) c_{\mathbf{k}\beta}$ and the nonzero components of ${\cal H}^{\alpha\beta}_{\rm tri-ka}$ are given explicitly in the Appendix, where $\alpha,\beta=0,1,\cdots,15$. The gap between the lowest two bands does not close as we change $R$ from $0.0$ to $1.0$ and Fig.~\ref{LatticeBandKagome} (b), (c) and (d) show the band structures at $R=0.0$, $0.5$ and $1.0$ as examples. Fig.~\ref{FlatBerryKagome} shows the band gap and band width as functions of $R$ as well as the Berry curvature at $R=0.0$, $0.5$ and $1.0$. We see that, similarly to the square-checkerboard lattice model, the Berry curvature changes significantly even though the energy dispersion remains quite flat at all $R$. 

\begin{figure}
\includegraphics[width=0.48\textwidth]{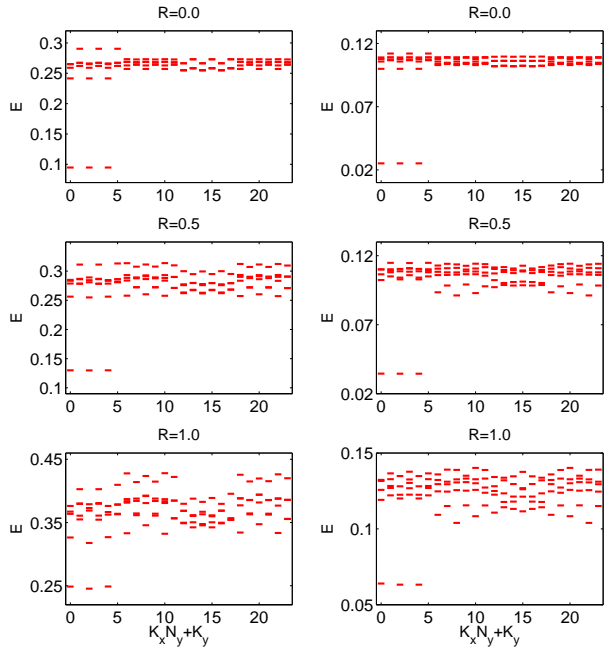}
\caption{Energy spectra at filling $1/3$ ($N=8$, $N_x=4$, $N_y=6$) for the square-checkerboard (left panels) and triangular-kagome (right panels) models at $R=0.0$, $0.5$ and $1.0$ (top to bottom). There are 3 quasidegenerate states at $(K_x,K_y)=(0,0)$, $(0,2)$ and $(0,4)$.}
\label{OneThirdGround}
\end{figure}

\begin{figure}
\includegraphics[width=0.48\textwidth]{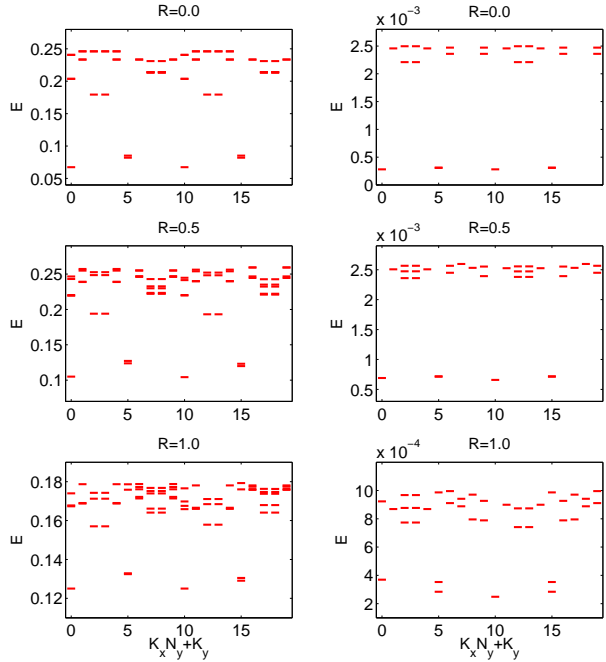}
\caption{Energy spectra at filling $1/2$ ($N=10$, $N_x=4$, $N_y=5$) for the square-checkerboard (left panels) and triangular-kagome (right panels) models at $R=0.0$, $0.5$ and $1.0$ (top to bottom). There are 6 quasidegenerate states: one each at $(K_x,K_y)=(0,0)$ or $(2,0)$, and two for $(1,0)$ or $(3,0)$.}
\label{OneHalfGround}
\end{figure}

\begin{figure}
\includegraphics[width=0.48\textwidth]{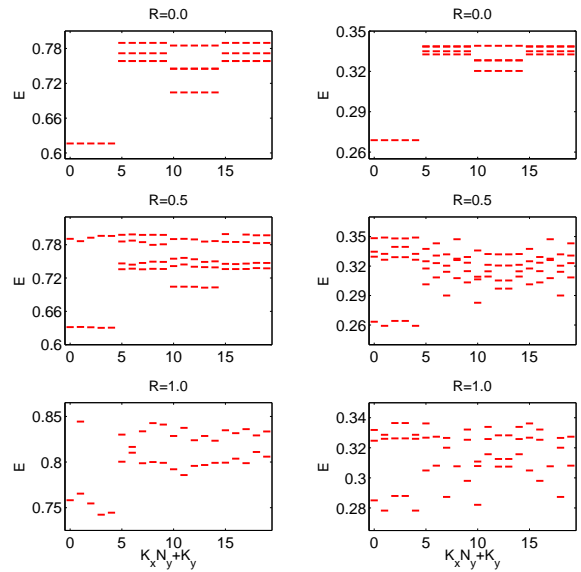} 
\caption{Energy spectra at filling $2/5$ ($N=8$, $N_x=4$, $N_y=5$) for the square-checkerboard (left panels) and triangular-kagome (right panels) models at $R=0.0$, $0.5$ and $1.0$. There are 5 quasi-degenerate states in each panel at $(K_x,K_y)=(0,0), (0,1), (0,2), (0,3)$ and $(0,4)$.}
\label{TwoFifthGround1}
\end{figure}

\begin{figure}
\includegraphics[width=0.48\textwidth]{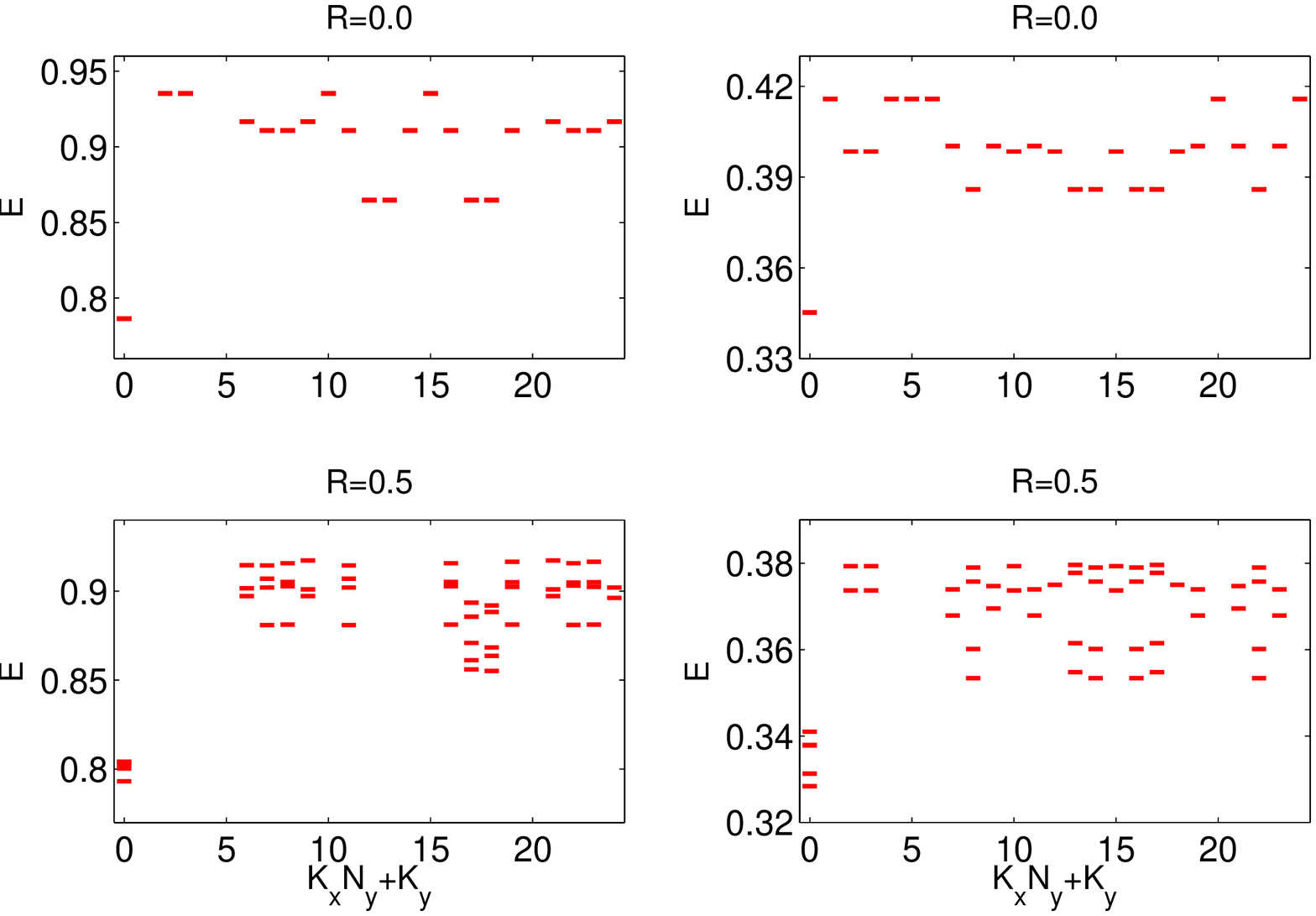}
\includegraphics[width=0.48\textwidth]{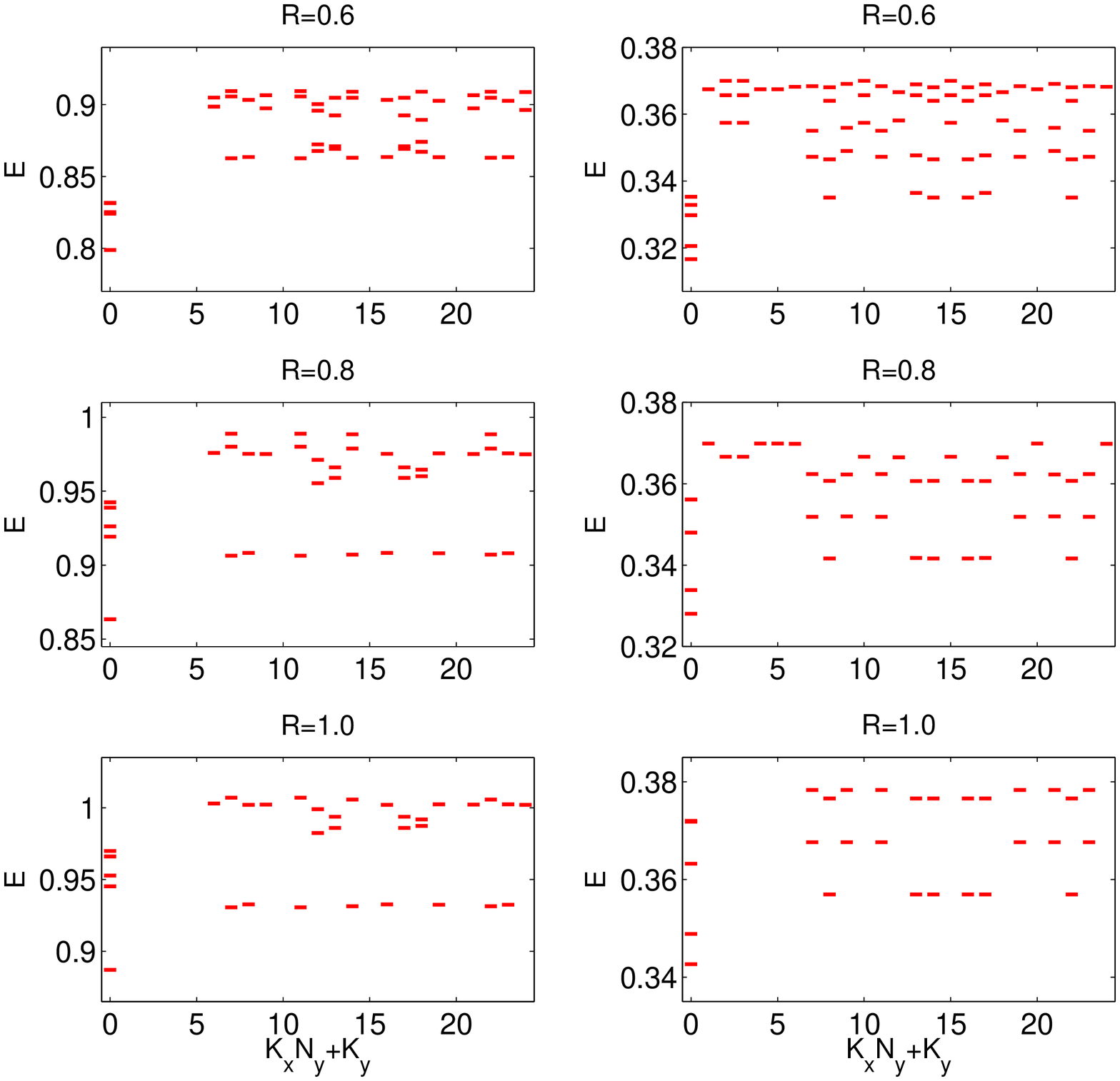}
\caption{Energy spectra at filling $2/5$ ($N=10$, $N_x=5$, $N_y=5$) for the square-checkerboard (left panels) and triangular-kagome (right panels) models at $R=0.0$, $0.5$, $0.6$, $0.8$ and $1.0$ (top to bottom). The 5 quasidegenerate states all appear in the $(K_x,K_y)=(0,0)$ sector and interaction causes splittings.}
\label{TwoFifthGround2}
\end{figure}

\begin{figure}
\includegraphics[width=0.48\textwidth]{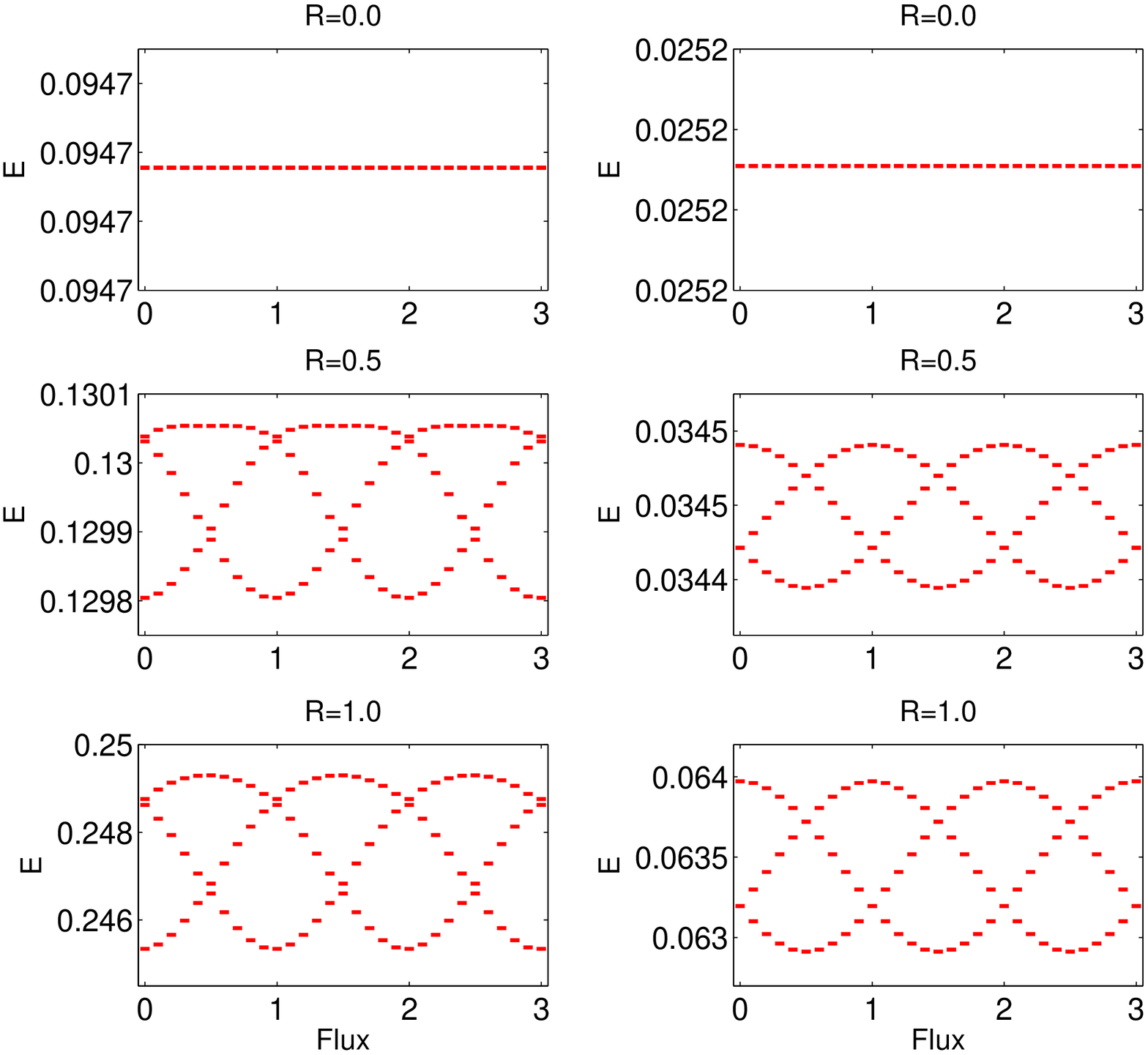}
\caption{Evolution of the $1/3$ ground states for square-checkerboard (left panels) and triangular-kagome (right panels) models shown in Fig.~\ref{OneThirdGround} upon flux insertion in the $y$-direction. The quasidegenerate ground states are separated from the excited state at each point. Note that at $R=0.0$, the states are perfectly degenerate at each flux value and there is no obvious spectral flow.}
\label{OneThirdInsert}
\end{figure}

\begin{figure}
\includegraphics[width=0.48\textwidth]{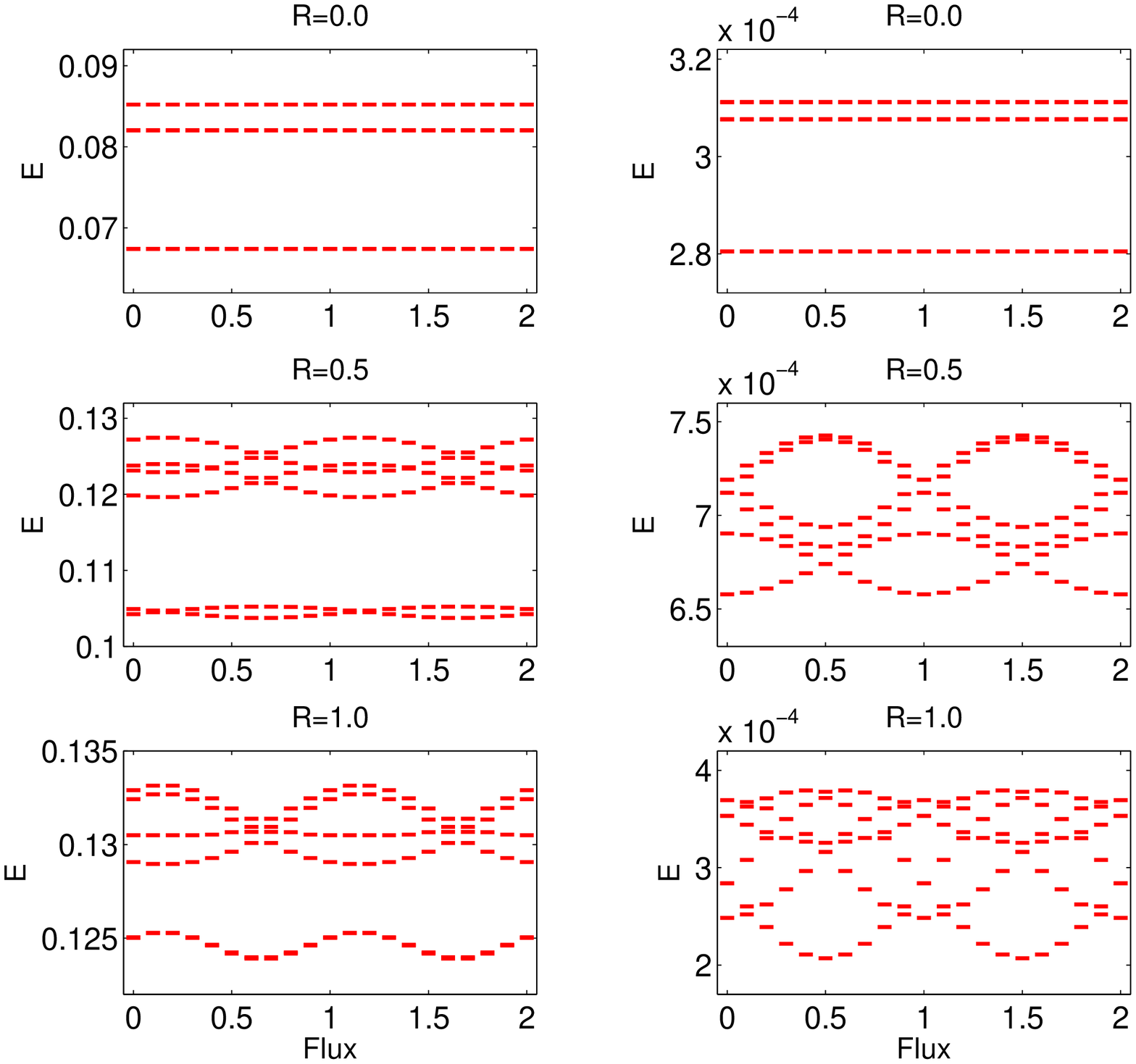}    
\caption{Evolution of the $1/2$ ground states for square-checkerboard (left panels) and triangular-kagome (right panels) models shown in Fig.~\ref{OneHalfGround} upon flux insertion in the $x$-direction. The quasidegenerate ground states are separated from the excited state at each point. Note that at $R=0.0$, the states are perfectly degenerate at each flux value and there is no obvious spectral flow.}
\label{OneHalfInsert}
\end{figure}

\begin{figure}
\includegraphics[width=0.48\textwidth]{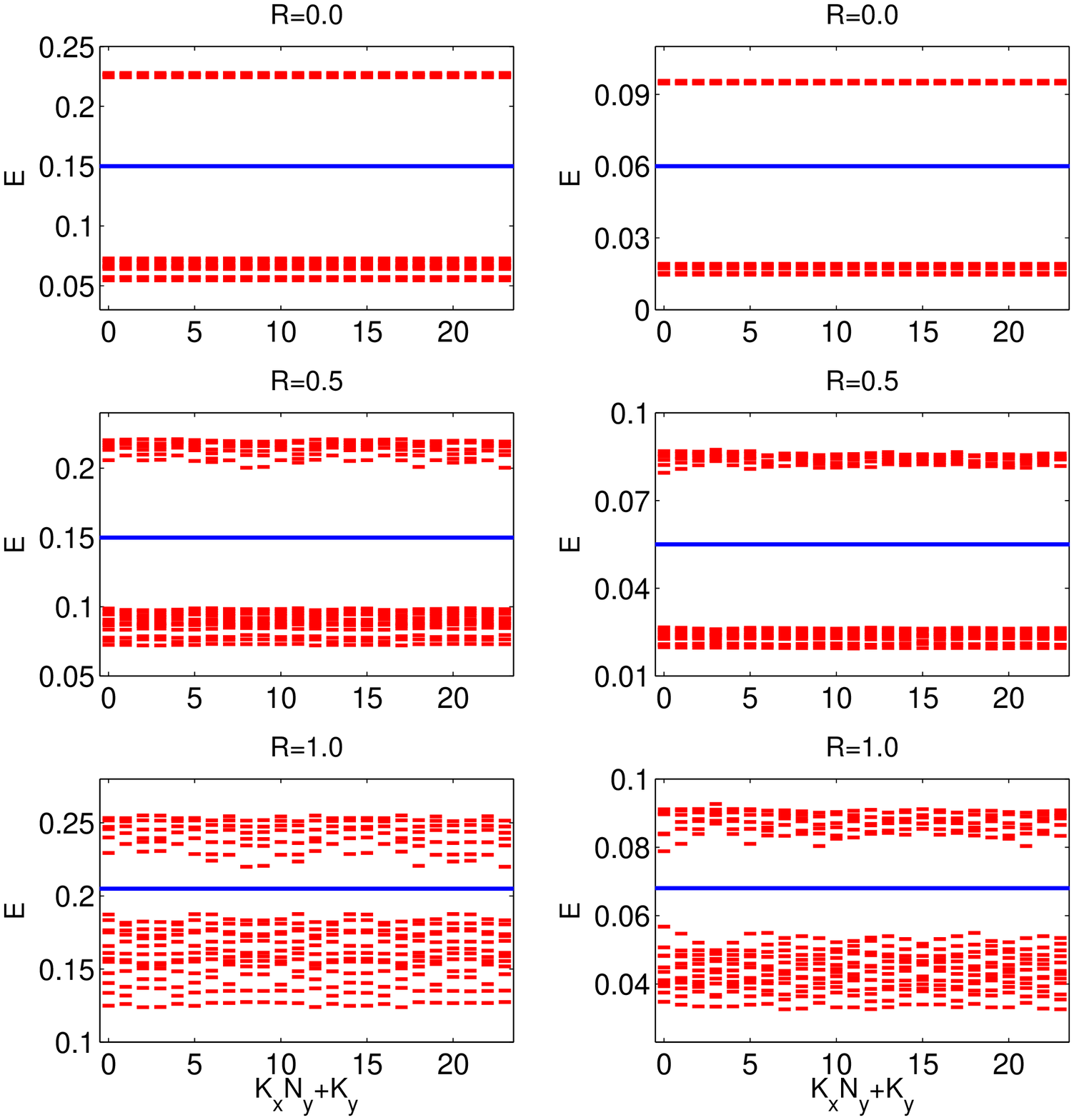}
\caption{Quasihole spectra at $1/3$ filling ($N=7$, $N_x=4$, $N_y=6$) for the square-checkerboard (left panels) and triangular-kagome (right panels) models at $R=0.0$, $0.5$ and $1.0$ (top to bottom). There are $12$ states in the low-energy manifold (below the blue lines) in each momentum sector.}
\label{OneThirdHole}
\end{figure}

\begin{figure}
\includegraphics[width=0.48\textwidth]{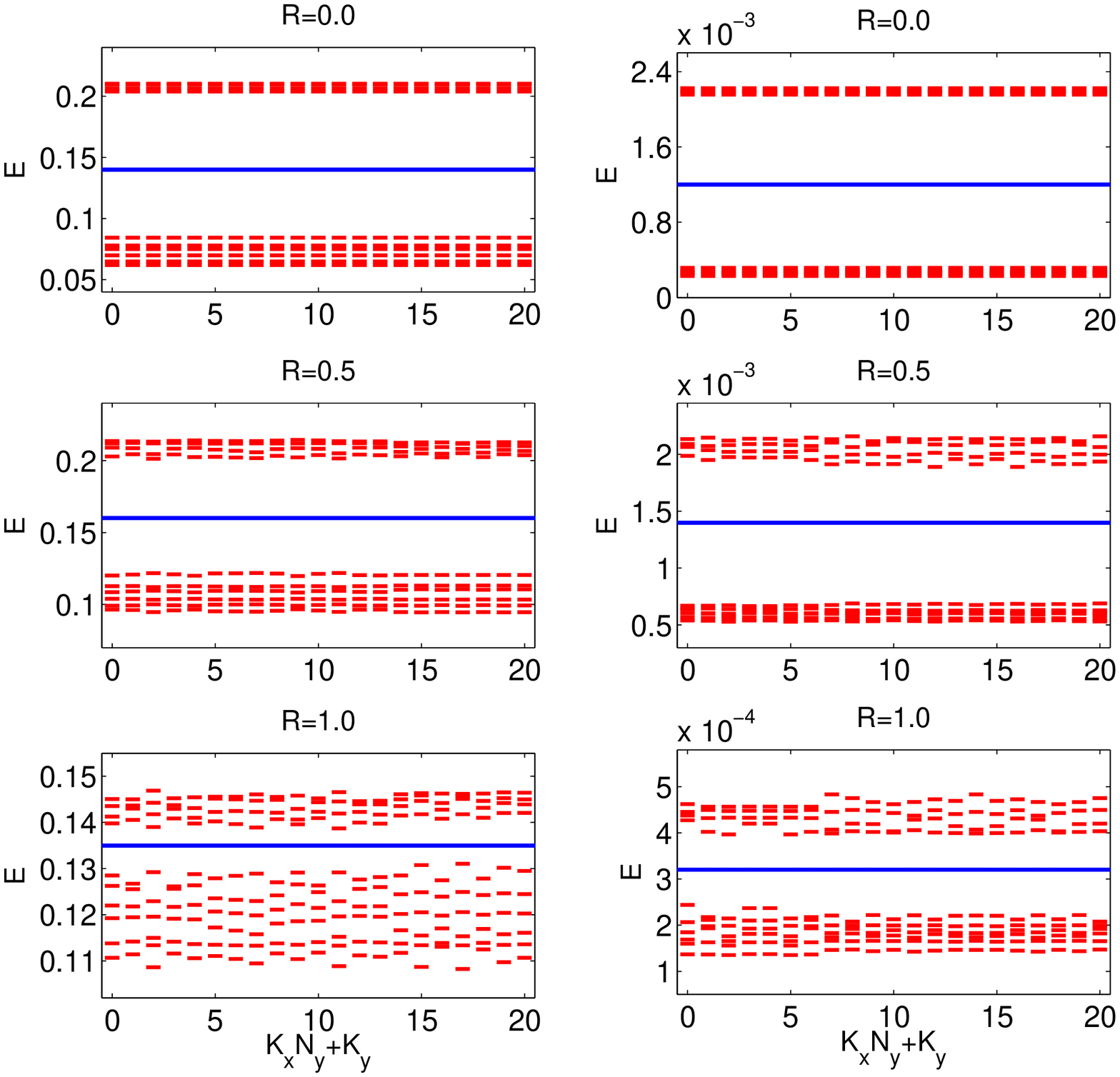}          
\caption{Quasihole spectra at filling $1/2$ ($N=10$, $N_x=3$, $N_y=7$) for the square-checkerboard (left panels) and triangular-kagome (right panels) models at $R=0.0$, $0.5$ and $1.0$ (top to bottom). There are $6$ states in the low-energy manifold (below the blue lines) in each momentum sector.}
\label{OneHalfHole}
\end{figure}

\begin{figure}                 
\includegraphics[width=0.48\textwidth]{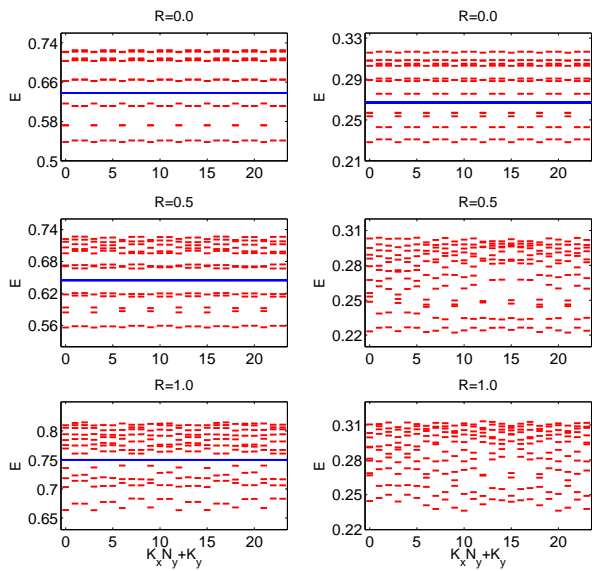}
\caption{Quasiparticle spectra at filling $1/3$ ($N=9$, $N_x=4$, $N_y=6$) for the square-checkerboard (left panels) and triangular-kagome (right panels) models at $R=0.0$, $0.5$ and $1$ (top to bottom). The number of states below the blue lines obey the FQH to FCI mapping in Eq.~(\ref{FoldRule}).}
\label{OneThirdParticle}
\end{figure}

\begin{figure}
\includegraphics[width=0.35\textwidth,height=0.22\textwidth]{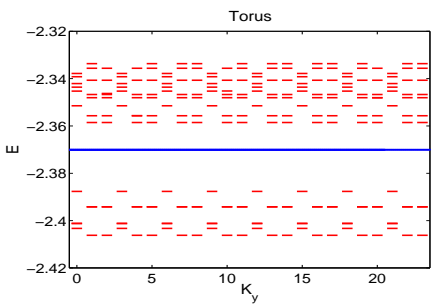}
\caption{Quasiparticle spectra at filling $1/3$ (9 particle and 24 fluxes) on torus with Coulomb interaction.}
\label{OneThirdTorus}
\end{figure}

\begin{figure}
\includegraphics[width=0.48\textwidth]{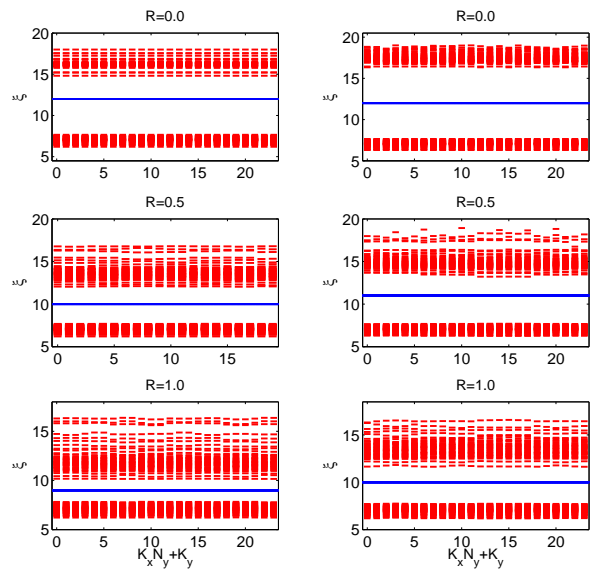}
\caption{Particle entanglement spectra at $1/3$ filling for the square-checkerboard (left panels) and triangular-kagome (right panels) models at $R=0$, $0.5$ and $1$ (top to bottom). The number of states in the low-entanglement-energy mainfold indicated by the blue lines are $46$ states in the $K_y=0$, $3$ momentum sectors, and $45$ states in other sectors.}
\label{OneThirdPES}
\end{figure}

\begin{figure}
\includegraphics[width=0.48\textwidth]{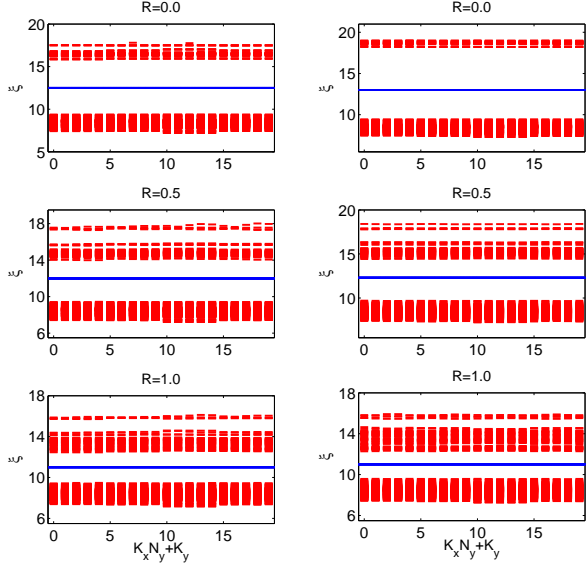}
\caption{Particle entanglement spectra at $1/2$ filling for the square-checkerboard (left panels) and triangular-kagome (right panels) model at $R=0$, $0.5$ and $1$ (top to bottom). The number of states in the low-entanglement-energy mainfold indicated by the blue lines are $200$, $196$, $201$ and $196$ states in the $K_x=0$, $1$, $2$, and $3$ momentum sectors, respectively.}
\label{OneHalfPES}
\end{figure}

\begin{figure}
\includegraphics[width=0.35\textwidth]{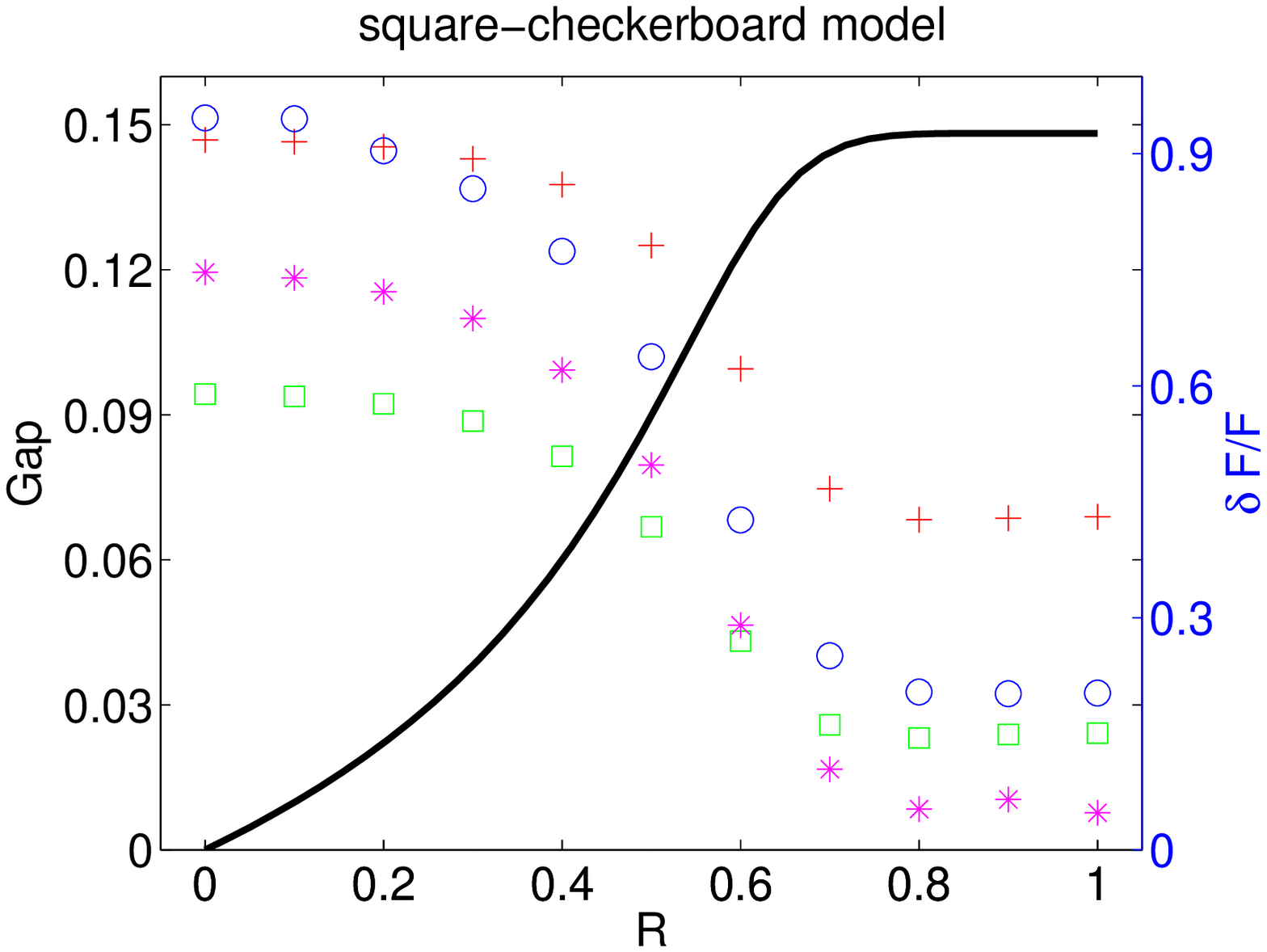}
\includegraphics[width=0.35\textwidth]{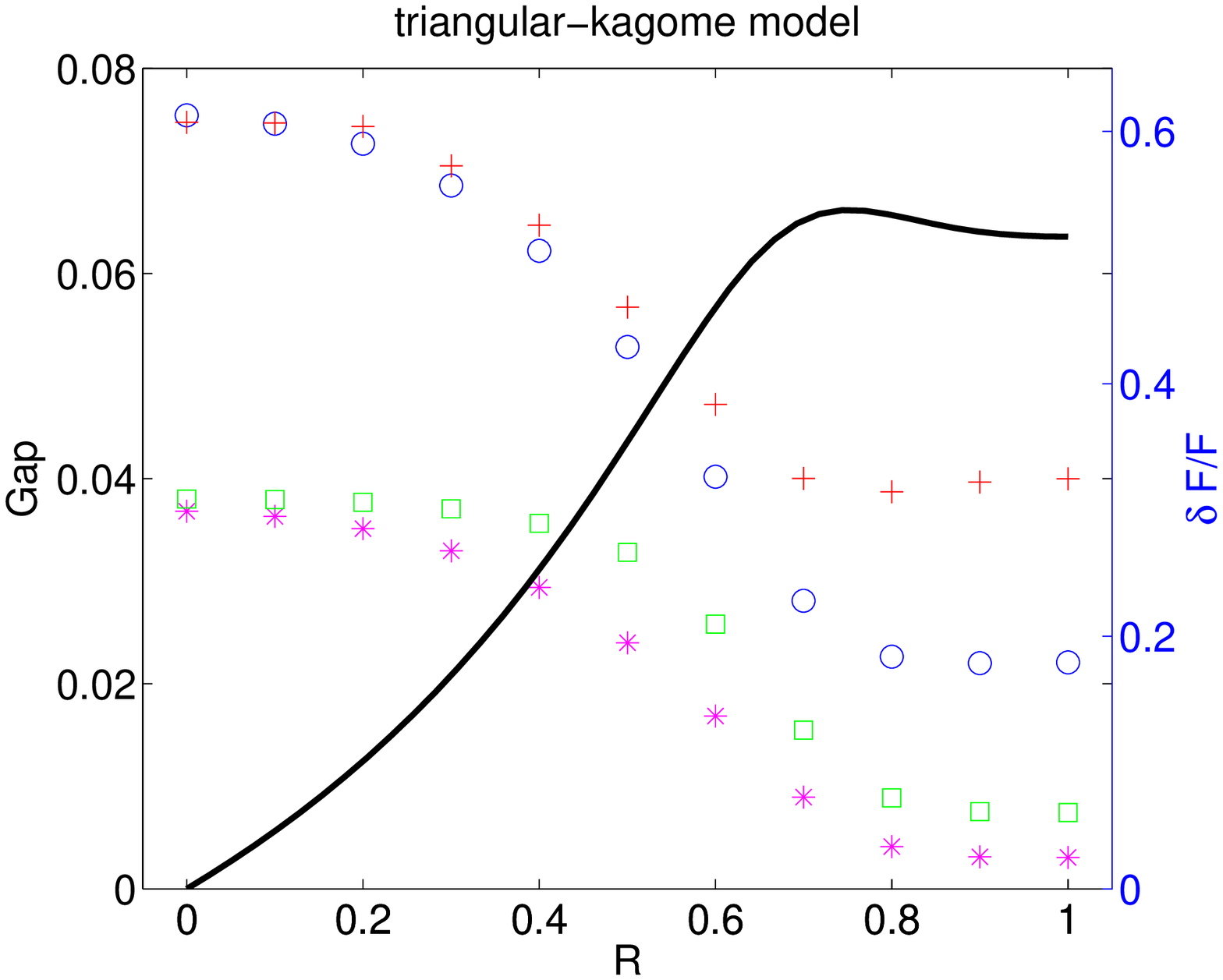}
\caption{The plus signs, circles, squares and asterisks show the gaps in the spectra of the $1/3$ ground states, $1/3$ quasihole states, $1/2$ ground states, and $1/2$ quasihole states at several $R$ for the square-checkerboard model (upper panel) and triangular-kagome model (lower panel). The continuous line shows the deviation of the Berry curvature $\delta \mathcal{F}$ (normalized by the average value $\bar{\mathcal{F}}$) as a function of $R$.}
\label{GapBerryRatio}
\end{figure}

\section{Fractional Quantum Hall Effect} 

Having shown that the $C=1$ bands of the Chern insulators are adiabatically connected to the Landau bands of Hofstadter lattices, we proceed to FQH states in these systems. As mentioned above, a Hofstadter lattice with a sufficiently small flux per plaquette simulates the continuum, and thus displays all the standard FQH states. Given that a Hofstadter insulator can trivially be converted into a Chern insulator, it follows, as a matter of principle, that all the FQH physics is also possible for Chern insulators, provided one chooses a sufficiently detailed unit cell. In this section, we will ask if the FCI states in the checkerboard and the kagome lattices can be understood as adiabatic evolutions of the corresponding states in Hofstadter lattices. Our exact diagonalization results below show that to be the case for the $1/3$ Laughlin state~\cite{Laughlin} and the $1/2$ Moore-Read state~\cite{MRPfaffian} in both models. However, the evidence for the $2/5$ Jain state is inconclusive. We stress that the FCI states at all these fractions had been established previously; our aim here is to show that they are adiabatic evolutions of the familiar FQH states at these fractions, thus establishing that the two are essentially the same.

\subsection{Model}

Our calculations are performed on lattices with periodic boundary conditions in the $x$ and $y$  directions with lengths $L_x$ and $L_y$. The number of particles and the number of magnetic unit cells in the $x$- and $y$-directions are denoted by $N$, $N_x$ and $N_y$. The adiabatic continuity between the lowest Landau level of the Hofstadter band and the Chern band clarifies that the filling factor is to be defined as $\nu=N/(N_xN_y)$. Following Ref.~[\onlinecite{Regnault}], we project out all high-energy bands to reduce the size of the Hilbert space, which is analogous to the lowest Landau level approximation routinely made in studies of the FQH states. We assume that the electrons are fully spin polarized. We also set the width of the lowest band to zero to eliminate the effect of the band curvature, which is a good approximation when the interaction energy is large compared to the bandwidth of the lowest band but small compared to the gap separating it to the first excited band. 

A short range interaction is known to produce FQH states at $n/(2n\pm 1)$ in a continuum Landau level, and, in particular, the 1/3 Laughlin state is the exact ground state of a short range interaction potential. For the FCI states at 1/3 and 2/5, we use a two-body interaction Hamiltonian 
\begin{equation}
H_2 = \sum_{[i,j]} U_{ij} {\hat n}_i {\hat n}_j 
\end{equation}
Here ${\hat n}_i=c_i^{\dagger} c_i$ is the particle number operator on site $i$. In order to make sure that the interaction has a nontrivial effect, we must make it sufficiently long ranged that it survives in the checkerboard limit. We will arrange it so that it becomes a nearest neighbor interaction in that limit. Specifically, we choose $U_{ij}=0.5/r_{ij}^2$ if the distance $r_{ij}$ between the sites $i$ and $j$ is smaller than or equal to a cutoff distance $r_c$, and $U_{ij}=0$ otherwise. For the square-checkerboard and triangular-kagome models, $r_c$'s are chosen to be $\sqrt{2}/2$ and $1/2$, respectively. All distances here and below are quoted in units of the lattice constant of the magnetic unit cell. We have also considered a truncated $1/r$ interaction and found that the results are consistent with our conclusions below; for simplicity, we will only show results for the truncated $1/r^2$ interaction.

For the FCI states at 1/2 we use a three-body interaction Hamiltonian
\begin{eqnarray}
H_3 = \sum_{[i,j,k]} V_{ijk} {\hat n}_i {\hat n}_j {\hat n}_k
\end{eqnarray}
where we choose $V_{ijk}$ to be $0.5/(r_{ij}r_{jk}r_{ki})^2$ if the distances $r_{ij}, r_{jk}, r_{ki}$ satisfy the cutoff conditions given below and $0$ otherwise, such that they become the nearest neighbor three body interaction in the checkerboard and kagome models. For the square-checkerboard lattice model, the condition is that $r_{ij,jk,ki} \le 1$ and $r_{ij} r_{jk} r_{ki} \le 1/2$. For the triangular-kagome lattice model, the condition is that $r_{ij,jk,ki} \le 1/8$ and $r_{ij} r_{jk} r_{ki} \le 1/2$.

\subsection{Exact Diagonalization}

We first calculate the eigenstates $u^n({\mathbf k})$ of the single-particle Hamiltonians ${\cal H}^{\alpha\beta}_{\rm sq-cb}$ or ${\cal H}^{\alpha\beta}_{\rm tri-ka}$. Then the Hamiltonians $H_{\rm sq-cb}$ or $H_{\rm tri-ka}$ are transformed to diagonal form by defining $c_{\mathbf{k}\alpha}= \sum_n u^n_{\alpha}({\mathbf k}) \gamma_{\mathbf{k}n}$, where $\gamma_{\mathbf{k}n}$ is the creation operators for a particle of momentum $\mathbf{k}$ in the $n$-th band. A many-body Slater basis state in the lowest band is given by $\gamma^\dagger_{\mathbf{k}_1,0} \gamma^\dagger_{\mathbf{k}_2,0} \ldots \gamma^\dagger_{\mathbf{k}_N,0} |0\rangle$, which has total momentum $\mathbf{k}_1+\mathbf{k}_2+\ldots+\mathbf{k}_N$. Since the Hamiltonians $H_2$ and $H_3$ commute with the translation operators in the $x$ and $y$ directions, they are block diagonal in the many-body basis. We decompose the Hilbert space into different sectors indexed by momentum quantum numbers $(K_x, K_y)$, which are the sum of the momentum all $N$ particles modulo $(N_x, N_y)$ in units of ($2\pi/L_x, 2\pi/L_y$). To calculate the many-body matrix elements of $H_2$ and $H_3$, we transform the Hamiltonians to momentum space in which they are expressed using $c^\dagger_{\mathbf{k}\alpha}$ and $c_{\mathbf{k}\alpha}$. The many-body Slater basis are defined only using the operators $\gamma^\dagger_{\mathbf{k},0}$, so the operator $c_{\mathbf{k}\alpha}$ is replaced by $u_{\alpha}({\mathbf k})\gamma_{\mathbf{k},0}$ when acting on these basis states.

\subsection{Ground States}

Fig.~\ref{OneThirdGround} and Fig.~\ref{OneHalfGround} shows the energy spectra at $1/3$ ($N=8$, $N_x=4$, and $N_y=6$) and $1/2$ ($N=10$, $N_x=4$ and $N_y=5$) fillings at $R=0.0$, $0.5$ and $1.0$. For $1/3$ filling, we observe $3$ quasidegenerate states at $(K_x,K_y)=(0,0)$, $(0,2)$ and $(0,4)$, while $6$ quasidegenerate states are found at $1/2$ filling: one for $(K_x,K_y)=(0,0)$ or $(2,0)$ and two for $(1,0)$ or $(3,0)$. The gap does not close as $R$ is increased from 0 to 1, as shown in Fig.~\ref{GapBerryRatio}, thus establishing an adiabatic continuity.

In Fig.~\ref{TwoFifthGround1} and Fig.~\ref{TwoFifthGround2}, we show the energy spectra of $H_2$ at filling factor $2/5$ with $N=8$, $N_x=4$, $N_y=5$ and $N=10$, $N_x=5$, $N_y=5$, respectively. We find that the $2/5$ states only show adiabatic continuity for the square-checkerboard model with $N=8$. For the triangular-kagome model with $N=8$ and for both models with $N=10$, however, the gap closes during the evolution. One may attribute the gap closing for the $N=10$ systems in the square-checkerboard model to a combination of the small gap and the fact that all ``ground states" occur at the same momenta and therefore are susceptible to significant mixing in finite systems; a study of larger systems will be necessary to clarify the fate of the 2/5 state in the square-checkerboard model.

The earlier work on FCI states at $1/3$, $1/2$ and $2/5$ in the checkerboard and kagome lattices used the degeneracy and momenta of the quasidegenerate ground states as criteria for identifying them with FQH-like states; these quantities are the same as the known degeneracy and momenta of the FQH ground states in the torus geometry for the same aspect ratio, and can be determined using root partitions and certain folding rules given by Bernevig and Regnault \cite{Bernevig}. The folding rule relates the degeneracy ${\cal N}_{\rm FQH}\left(K_x , K_y \right)$ of low energy FQH states in the $\left(K_x , K_y \right)$ momentum sector and the approximate degeneracy ${\cal N}_{\rm FCI}\left(K_x , K_y \right)$ for the FCI case via the following equation
\begin{eqnarray}
{\cal N}_{\rm FCI}\left(K_x, K_y\right)=\sum_{K_x',K_y'=0}^{N-1} \delta_{K_x'{\rm mod}N_{x0},K_x} \nonumber \\
\times \delta_{K_y'{\rm mod}N_{y0},K_y} \frac{N_{x0}N_{y0}}{N_0} {\cal N}_{\rm FQH} \left(K_x',K_y'\right) 
\label{FoldRule}
\end{eqnarray}
where $N_{x0}={\rm GCD}(N,N_x)$, $N_{y0}={\rm GCD}(N,N_y)$ and $N_0={\rm GCD}(N,N_xN_y)$ (${\rm GCD}$ denotes the greatest common divisor). For the $1/3$ Laughlin state and the $1/2$ Moore-Read state, the degeneracy of ground states and quasihole states (discussed below) ${\cal N}_{\rm FQH}\left(K_x , K_y \right)$ can be obtained using a generalized Pauli principle~\cite{Jack} and the many- body translational symmetry~\cite{Haldane2}. For general composite fermion states, the usage of the generalized Pauli principle is limited, but we can still directly compare the energy spectrum of a FQH system on torus and its counterpart in a FCI to check the validity of Eq.~(\ref{FoldRule}). For our considerations, however, an {\em a priori} knowledge of the counting is not necessary, as we directly establish adiabatic continuity with the reference FQH state in the Hofstadter limit. This becomes important when the generalized Pauli principle does not apply, {\em e.g.} for the quasiparticle spectra (below).

Further proof that the state has a fractional Hall conductance (or a fractional Chern number) can be demonstrated by looking at the evolution of the quasidegenerate ground states upon flux insertion along the $x$ or $y$ direction. The effect of inserting a flux $\Phi$ in either of the two directions is implemented by letting the single-particle momenta $k_{x,y} \rightarrow k_{x,y} + \Phi$. For non-FQH states, a state will come back to itself after one flux insertion, whereas a FQH state returns to the original state only after insertion of several flux quanta. Fig.~\ref{OneThirdInsert} show that at 1/3, one of the quasidegenerate ground state evolves into a second state after one flux quantum, and into a third after two flux quanta, before returning to the original state. This demonstrates a Hall conductance of $1/3$. Note that there is no level crossing with higher energy states. Similarly, in Fig.~\ref{OneHalfInsert}, the ground states only evolve back to themselves, without crossing higher-energy levels, after inserting two flux quanta, which reveals the $1/2$ Hall conductance. 

\subsection{Quasiholes and quasiparticles}

A FQH state is characterized not only by its ground state but also by the nature of its quasiholes and quasiparticles, in particular the number of quasidegenerate states when one or several quasiholes or quasiparticles are created. For FQH states, the composite fermion (CF) theory has been shown (in the spherical geometry) to give a complete account of states containing quasiparticles or quasiholes for the fractions of the form $n/(2pn\pm 1)$, such as the number of quasi-degenerate states, their quantum numbers (orbital angular momenta for the spherical geometry), and their wave functions;~\cite{CFBook} this demonstrates that the quasiparticles are composite fermions in a nearly empty $\Lambda$ level and quasiholes are missing composite fermions from an almost full $\Lambda$ level. Unfortunately, formulation of the CF theory in the torus geometry is not yet available, but we can take the solution in the Hofstadter limit as our definition of the quasihole or quasiparticle spectrum (provided a low energy band can be clearly identified). For quasiholes, the number of states in each momentum sector can also be obtained using the generalized Pauli principle and the folding rules.~\cite{Bernevig} 

In Fig.~\ref{OneThirdHole} and Fig.~\ref{OneHalfHole}, the quasihole spectra with $N=7$, $N_x=4$ and $N_y=6$ and $N=10$, $N_x=3$ and $N_y=7$ are presented, which correspond to a $\nu=1/3$ and $1/2$ states with three and two quasiholes, respectively. The principal observation is that the gap between the low energy quasihole manifold and the higher energy states does not close as $R$ is increased from 0 to 1, as shown by the circles and asterisks in Fig.~\ref{GapBerryRatio}. 

In Fig.~\ref{OneThirdParticle}, we show energy spectra of $H_2$ with $N=9$, $N_x=4$ and $N_y=6$ which correspond to the $1/3$ state with $3$ quasiparticles ({\em i.e.}, three composite fermions in the second $\Lambda$ level). Here, it is not clear, even at $R=0$, how to identify the quasiparticle band. For this purpose, we show in Fig.~\ref{OneThirdTorus} the energy spectra for the corresponding FQH state (9 electrons on a torus interacting via the Coulomb interaction in the presence of 24 flux quanta) on a torus. This system has a well defined quasiparticle band, which allows us to also identify the quasiparticle bands at $R=0$ in the current problem as well, as marked by the blue lines. This band is seen to evolve continuously, without gap closing, in the square-checkerboard model, but not in the triangular-kagome limit. The gaps in the quasiparticle spectra are not as clear as those in the quasihole cases; as a confirmation of our assignment of the quasiparticle bands, we also studied flux insertion in these systems and found that the states marked under the blue lines do not mix with higher-energy states above the lines. 

The fact that the quasiparticle band is not very well defined is already an indication that the 2/5 state will be either weak or absent in the checkerboard lattice and absent in the kagome lattice. The large bandwidth of the quasiparticle band implies substantial residual interactions between composite fermions in the second $\Lambda$ level, which can weaken or destroy the two-filled-$\Lambda$-level 2/5 state. The $n/(2pn\pm 1)$ states, which are the prominent FQH states in the lowest Landau level of the continuum, are even more unlikely to occur in Chern bands for $n\geq 3$.

\subsection{Particle entanglement spectra}

The entanglement spectrum \cite{LiHaldane} has been used to probe the topological properties of many FQH states \cite{TwoFifthES,Sterdyniak,Papic}. For the torus geometry used here, the particle entanglement spectrum (PES) \cite{Sterdyniak} has proven particularly useful. Given $d$ (quasi-)degenerate ground states $\{|\psi_i\rangle\}$, the density matrix is defined as $\rho=d^{-1}\sum_{i=1}^{d}|\psi_i\rangle \langle\psi_i|$. We make a cut in the particle space by dividing the $N$ particles into two groups $A$ and $B$ with $N_A$ and $N_B$ particles. The reduced density matrix $\rho_A={\rm Tr}_B \rho$ is obtained by tracing out the particles in $B$. The translational symmetries along the $x$ and $y$ directions are preserved in this process, so we can plot the eigenvalues $\exp(-\xi)$ ($\xi$ is usually called the entanglement energy) of $\rho_A$ versus the momenta of their corresponding eigenstates. As previously found \cite{Regnault,Bernevig,Wu1}, the numbers of low-lying levels in the PES are determined by the numbers of quasihole states that $N_A$ particles can form on an $N_x \times N_y$ lattice. There are also levels at higher entanglement energies separated from the low-lying universal ones by ``entanglement gaps." It is further demonstrated that the PES can differentiate FQH states from charge density wave states which occur in the thin torus limit \cite{ThinTorus}. 

The PES are presented in Fig.~\ref{OneThirdPES} and Fig.~\ref{OneHalfPES} for $1/3$ and $1/2$ fillings. We trace out $5$ and $6$ particles in these two cases, respectively. For the $1/3$ PES, the low energy band (below the blue lines) consists of $46$ states at each momentum in the $K_y=0$, $3$ momentum sectors and $45$ states in other sectors. For the $1/2$ PES, the low energy band has $200$, $196$, $201$ and $196$ states in the $K_x=0$, $1$, $2$ and $3$ momentum sectors, respectively. These numbers agree with theoretical predictions \cite{Bernevig} and the entanglement gap does not close for any value of $R$ between $0$ and $1$.

\subsection{Role of the Berry curvature}

We now ask what weakens or destroys FQH states as the Hofstadter lattices evolve to Chern insulators. It is clear that the nonflatness of the bands is not relevant here, both because the lowest bands are quite flat over the entire evolution as shown in panels (a) of Fig.~\ref{FlatBerryChecker} and Fig.~\ref{FlatBerryKagome}, and because we have set the bands to be strictly flat by hand. We believe that the relevant quantity in this respect is the nonuniformity (in momentum space) of the Berry curvature. The adiabatic path between a FQH insulator and a FCI offers a natural way to explore the role of the distribution of the Berry curvature in momentum space.  

Panels (b) of Fig.~\ref{FlatBerryChecker} and Fig.~\ref{FlatBerryKagome} show the distribution of the Berry curvature ($\mathcal{F}$) as a function of $R$ along certain lines in the Brillouin zone for our models. In both cases, the Berry curvature of the lowest band is flat at $R=0$, as expected for a Landau level. However, in the the square-checkerboard model, the value of $\mathcal{F}$ reduces near the $\Gamma$ and $M$ points, and has a peak at the $X$ point as $R$ increases. In the triangular-kagome model, a peak of $\mathcal{F}$ emerges at the $K$ point while its value near the $\Gamma$ point goes to zero. Although the integrated Berry curvature in the whole Brillouin zone remains constant ($2\pi$ times the Chern number), its fluctuations become nearly as large as its mean value as $R$ approaches unity.

This change in the distribution of the Berry curvature has a direct correlation with the many body properties.  As a quantitative measure of the deviation of the Berry curvature from its average value $\bar{\mathcal{F}}$, we define the standard deviation of Berry curvature as
\begin{eqnarray}
\mathcal{\delta F} &=& \sqrt{ \frac{1}{\mathcal{A}} \int_{\rm BZ} d^2 k (\mathcal{F}(\mathbf{k}) - \bar{\mathcal{F}})^2}
\end{eqnarray}
with
\begin{eqnarray}
\bar{\mathcal{F}} &=& \frac{1}{\mathcal{A}} \int_{\rm BZ} d^2 k \mathcal{F}(\mathbf{k})
\end{eqnarray}
where $\mathcal{A}$ is the area of the Brillouin zone. We now argue that this quantity determines the robustness of the FQH states. We see from Fig.~\ref{GapBerryRatio} that the spectral gaps for the ground and quasihole states at different fillings follow the same trend, changing rapidly for $R<0.7$ and then saturating at $R\sim 0.7$. These many-body gaps have a strong (anti)correlation with $\delta\mathcal{F}$, indicating its important role in the FCI states. While the topological properties remain intact over a wide range of $\delta \mathcal{F}$, an increase in $\delta \mathcal{F}$ reduces the size of the gap and thus the robustness of the FCI states. Essentially, the variations in ${\cal F}$ enhance the residual interactions between composite fermions, as we see from the increase in the bandwidths of the quasihole and quasiparticle states, which causes a weakening of the FCI states. However, interactions between composite fermions open up the possibility of new emergent structure. 

\section{Concluding Remarks} 

A remark on the form of the interaction is in order. By construction, the interaction becomes a nearest-neighbor interaction in the checkerboard and the kagome limits, independent of what short range decay is assumed. However, the robustness of the states in the Hofstadter limit depends on the form of the interaction. As mentioned previously, we find that using an exponent of 1 rather than 2 in $H_2$ does not change our results qualitatively for the 1/3 state, in the sense that adiabatic continuity can still be established and the gap decreases as the deviation of Berry curvature $\delta{\cal F}$ increases. On the contrary, if we choose the exponent in $H_3$ to be 1 rather than 2 then the $1/2$ FQH state in the Hofstadter lattice ({\em i.e.} $R=0$) is much weaker, with small spectral gaps and relatively large ground state splittings; in this case, the gap in the checkerboard limit is actually larger than that in the Hofstadter limit, so our conclusion that the gap decreases with increasing $\delta{\cal F}$ does not hold. This is physically understandable. The Moore-Read Pfaffian state is the exact zero-energy state of a short-range three-body interaction in the lowest Landau level, but adding longer-range components weakens, and even eliminates, this state~\cite{Wojs}. A smaller exponent in $H_3$ means a longer-range interaction in the Hofstadter limit, which renders the $1/2$ FQH states weaker.

The equivalence of the FQH and FCI states has been studied from the perspective of density algebra,~\cite{Sid,Goerbig} based on the observation that the commutators of momentum-space density operators have the same form in the long-wavelength (small-momentum) limit and for flat Berry curvature. This suggests the same low energy physics for the two problems. However, all momenta are relevant to the FCI states since the bands are exactly flat and the low-energy excitations may have large momenta. Very recently, Roy~\cite{Roy} demonstrated that the density algebra in a lattice model has the same form as that of the Landau level in continuum for all momenta if the Fubini-Study metric satisfies a certain condition. The Hofstadter models that we construct have nearly flat Berry curvature and thus the folding rule is almost exact. Our work shows that the low energy physics can evolve adiabatically as the Berry curvature changes, and thus provides justification for the assumption of flat Berry curvature in the aforementioned works.

We use the particle entanglement spectrum as a probe of the quasihole physics of our models. Ref.~\onlinecite{LiuBerg} obtains the orbital entanglement spectra~\cite{LiHaldane,Haque} for the kagome model at both $1/3$ and $1/2$ fillings, which reveal the edge modes of these states and provide further support for the adiabatic continuity between the FQH states in continuum and FCI states on lattice.
 
We do not find conclusive evidence for adiabatic continuity for the $2/5$ state in either model, which is due to absence of the $2/5$ FCI states. Recent papers have proposed that $2/5$ states can be obtained in the checkerboard model~\cite{Lauchli} and the kagome model~\cite{Liu} by either using tilted samples or fine tuning of parameters. We believe that these states are also adiabatically connected to the $2/5$ state in the Hofstadter lattice, but we have not confirmed this. Similarly, if the quasiparticle spectra of the Chern insulator models can be obtained from the FQH quasiparticle spectra on torus via the folding rule at $R=1$ after fine tuning of parameters, we do expect adiabatic continuity between the quasiparticle spectra at $R=0$ and $R=1$. 

In conclusion, we have shown, by studying the ground states, quasihole and quasiparticle states, and their particle entanglement spectra, that the integer and fractional states in the Hofstadter and Chern insulators are adiabatically connected. Our study reveals that the nonuniform distribution of the Berry curvature reduces the gap and increases the interaction strength between quasiparticles. In addition, our work shows how Chern insulators with arbitrarily uniform Berry curvature can be constructed by allowing more complex lattices, which should produce many other FCI states. Time-reversal-invariant fractional topological insulators can be constructed from the $p/(2p+1)$ states by introducing a spin, and are expected to be topologically stable for odd $p$.~\cite{Ferraro}

{\em Note added} -- At the time of preparing the first version of this manuscript, we became aware of a preprint \cite{Scaffidi} on a similar topic.

\section*{Acknowledgements}

We are indebted to the authors, especially N. Regnault, of the DiagHam package, and acknowledge financial support from the JQI-NSF-PFC (KS) and DOE Grant no. DE-SC0005042 (YHW and JKJ). YHW and JKJ thank the Joint Quantum Institute and the Condensed Matter Theory Center, University of Maryland, for their kind hospitality. We thank Research Computing and Cyberinfrastructure, a unit of Information Technology Services at The Pennsylvania State University, for providing high-performance computing resources and services used for the computations in this work.

\newpage
\pagebreak

\begin{appendix}
\begin{widetext}

\section{Matrix Elements of the Single-Particle Hamiltonians in the Momentum Space}

We first give the matrix elements of ${\cal H}_{\rm sq-cb}$, the single particle Hamiltonian for the square-checkerboard hybrid lattice model. The nonzero matrix elements are (dropping the subscript ``sq-cb" for simplicity)
\begin{eqnarray}
& {\cal H}^{01}=\exp(ik_x/4), \;\; {\cal H}^{03}=\exp(-i({k_x}/4+{3\pi}/80)), \;\; {\cal H}^{04}=\exp(i({k_x}/4+{3\pi}/80)),
\nonumber \\ 
& {\cal H}^{0,12}=\exp(-i{k_y}/4), \;\; {\cal H}^{12}=\exp(i({k_x}/4+{\pi}/16)), \;\; {\cal H}^{15}=\exp(i({k_y}/4+{\pi}/5)),
\nonumber \\
&  {\cal H}^{1,13}=\exp(-i({k_y}/4+{\pi}/16)), \;\; {\cal H}^{23}=\exp(i({k_x}/4+{3\pi}/40)), \;\; {\cal H}^{26} = \exp(i({k_x}/4+{37\pi}/80)), 
\nonumber \\
& {\cal H}^{2,14}=\exp(-i({k_y}/4+{3\pi}/40)), \;\; {\cal H}^{37}=\exp(i({k_y}/4+{39\pi}/40)), \;\; {\cal H}^{3,15}=\exp(-i({k_y}/4+{3\pi}/80)),
\nonumber \\ 
& {\cal H}^{45}=\exp(i({k_x}/4+{3\pi}/80)), \;\; {\cal H}^{47}=\exp(-i({k_x}/4+{39\pi}/40)), \;\; {\cal H}^{48}=\exp(i({k_y}/4+{3\pi}/40)), 
\nonumber \\ 
& {\cal H}^{56}=\exp(i({k_x}/4+{\pi}/5)), \;\; {\cal H}^{59}=\exp(i({k_y}/4+{19\pi}/80)), \;\; {\cal H}^{67}=\exp(i({k_x}/4+{37\pi}/80)), 
\nonumber \\ 
& {\cal H}^{6,10}=\exp(i({k_y}/4+{2\pi}/5)), \;\; {\cal H}^{7,11}=\exp(i({k_y}/4+{37\pi}/80)), \;\; {\cal H}^{89}=\exp(i({k_x}/4+{3\pi}/40)), 
\nonumber \\ 
& {\cal H}^{8,11}=\exp(-i({k_x}/4+{37\pi}/80)), \;\; {\cal H}^{8,12}=\exp(i({k_y}/4+{\pi}/16)), \;\; {\cal H}^{9,10}=\exp(i({k_x}/4+{19\pi}/80)), 
\nonumber \\
& {\cal H}^{9,13}=\exp(i({k_y}/4+{7\pi}/40)), \;\; {\cal H}^{10,11}=\exp(i({k_x}/4+{2\pi}/5)), \;\; {\cal H}^{10,14}=\exp(i({k_x}/4+{19\pi}/80)),  
\nonumber \\ 
& {\cal H}^{11,15}=\exp(i({k_y}/4+{\pi}/5)), \;\; {\cal H}^{12,13}=\exp(i({k_x}/4+{\pi}/16)), \;\; {\cal H}^{12,15}=\exp(-i({k_x}/4+{\pi}/5)), 
\nonumber \\
& {\cal H}^{13,14}=\exp(i({k_x}/4+{7\pi}/40)), \;\; {\cal H}^{14,15}=\exp(i({k_x}/4+{19\pi}/80)), \nonumber \\ 
& {\cal H}^{44}=(-2t_2)(\cos(k_x+7\pi/40)+\cos(k_y-47\pi/40))+2t_3(\cos(k_x-k_y+47\pi/20)+\cos(k_x+k_y)), 
\nonumber \\
& {\cal H}^{4,14}=(-t_1)(\exp(i(k_x/2+k_y/2+3\pi/10))+\exp(i(k_x/2-k_y/2+79\pi/40)), \nonumber \\
& + \exp(i(-k_x/2+k_y/2-3\pi/8))+\exp(i(-k_x/2-k_y/2+3\pi/10)), \nonumber \\
& {\cal H}^{14,14}=(-2t_2)(\cos(k_x+47\pi/40)+\cos(k_y-7\pi/40))+2t_3(\cos(k_x-k_y+47\pi/20)+\cos(k_x+k_y)).
\end{eqnarray}
where $t_1=1$, $t_2=1-\sqrt{2}/2$ and $t_3=(\sqrt{2}-1)/2$. We note that ${\cal H}^{44}$, ${\cal H}^{4,14}$ and ${\cal H}^{14,14}$ should be multiplied by $R$, whereas the remaining elements should be multiplied by $1-R$; these factors were omitted for notational ease. Other elements can be obtained by complex conjugation.   

Next we give the matrix elements of ${\cal H}_{\rm tri-ka}$, the single particle Hamiltonian for the triangular-kagome hybrid lattice model. The nonzero matrix elements are (dropping the subscript ``tri-ka" for simplicity)
\begin{eqnarray}
& {\cal H}^{01}=\exp(ik_x/4), \;\; {\cal H}^{03}=\exp(-ik_x/4), \;\; {\cal H}^{04}=\exp(ik_y/4) 
\nonumber \\ 
& {\cal H}^{07}=\exp(i(-k_x/4+k_y/4-{25\pi}/16)), \;\; {\cal H}^{0,12}=\exp(ik_y/4), \;\;  {\cal H}^{0,13}=\exp(i(k_x/4-k_y/4-{\pi}/16)),
\nonumber \\
& {\cal H}^{12}=\exp(ik_x/4), \;\; {\cal H}^{14}=\exp(i(-k_x/4+k_y/4+{\pi}/16)), \;\; {\cal H}^{15}=\exp(i(k_y/4+{\pi}/8)),
\nonumber \\ 
& {\cal H}^{1,13}=\exp(-i(k_y/4+{\pi}/8)), \;\; {\cal H}^{1,14}=\exp(i(k_x/4-k_y/4-3{\pi}/16))), \;\; {\cal H}^{23}=\exp(ik_x/4),
\nonumber \\ 
& {\cal H}^{25}=\exp(i(-k_x/4+k_y/4+{3\pi}/16)), \;\; {\cal H}^{26}=\exp(i(k_y/4+{\pi}/4)), \;\; {\cal H}^{2,14}=\exp(-i(k_y/4+{\pi}/4)),  
\nonumber \\ 
& {\cal H}^{2,15}=\exp(i(k_x/4-k_y/4-{5\pi}/16)), \;\; {\cal H}^{36}=\exp(i(-k_x/4+k_y/4+{5\pi}/16)), \;\; {\cal H}^{37}=\exp(i(k_y/4+{3\pi}/8)), 
\nonumber \\
& {\cal H}^{3,12}=\exp(i(k_x/4-k_y/4+{\pi}/16)), \;\; {\cal H}^{3,15}=\exp(-i(k_y/4+{3\pi}/8)), \;\; {\cal H}^{45}=\exp(ik_x/4),
\nonumber \\
& {\cal H}^{47}=\exp(-i(k_x/4+{3\pi}/2)), \;\; {\cal H}^{48}=\exp(ik_y/4), \;\; {\cal H}^{4,11}=\exp(i(-k_x/4+k_y/4-{17\pi}/16)), 
\nonumber \\
& {\cal H}^{56}=\exp(ik_x/4), \;\; {\cal H}^{58}=\exp(i(-k_x/4+k_y/4+{\pi}/16)), \;\; {\cal H}^{59}=\exp(i(k_y/4+{\pi}/8)),
\nonumber \\ 
& {\cal H}^{67}=\exp(ik_x/4), \;\; {\cal H}^{69}=\exp(i(-k_x/4+k_y/4+{3\pi}/16)), \;\; {\cal H}^{6,10}=\exp(i(k_y/4+{\pi}/4)), 
\nonumber \\
& {\cal H}^{7,10}=\exp(i(-k_x/4+k_y/4+{5\pi}/16)), \;\; {\cal H}^{7,11}=\exp(i(k_y/4+{3\pi}/8)), \;\; {\cal H}^{89}=\exp(ik_x/4),  
\nonumber \\
& {\cal H}^{8,11}=\exp(-i(k_x/4+\pi)), \;\; {\cal H}^{8,12}=\exp(ik_y/4), \;\; {\cal H}^{8,15}=\exp(i(-k_x/4+k_y/4-{9\pi}/16)),
\nonumber \\
& {\cal H}^{9,10}=\exp(ik_x/4), \;\; {\cal H}^{9,12}=\exp(i(-k_x/4+k_y/4+{\pi}/16)), \;\; {\cal H}^{9,13}=\exp(i(k_y/4+{\pi}/8)),
\nonumber \\
& {\cal H}^{10,11}=\exp(ik_x/4) \;\; {\cal H}^{10,13}=\exp(i(-k_x/4+k_y/4+{3\pi}/16)) \;\; {\cal H}^{10,14}=\exp(i(k_y/4+{\pi}/4))
\nonumber \\
& {\cal H}^{11,14}=\exp(i(-k_x/4+k_y/4+{5\pi}/16)), \;\; {\cal H}^{11,15}=\exp(i(k_y/4+{3\pi}/8)), \;\; {\cal H}^{12,13}=\exp(ik_x/4),
\nonumber \\
& {\cal H}^{12,15}=\exp(-i(k_y/4+{\pi}/2)), \;\; {\cal H}^{13,14}=\exp(ik_x/4), \;\; {\cal H}^{14,15}=\exp(ik_x/4),
\nonumber \\
& {\cal H}^{02}=-2(t_1\cos(k_x/2)+t_2\cos(-k_x/2+k_y)), \nonumber \\
& {\cal H}^{28}=-2(t_1\cos(-k_x/2+k_y/2)+t_2\cos(k_x/2+k_y/2)), \nonumber \\
& {\cal H}^{80}=-2(t_1\cos(k_y/2)+t_2\cos(k_x-k_y/2)).
\end{eqnarray}
where $t_1=1.0+0.28i$ and $t_2=-0.3-0.2i$. The elements ${\cal H}^{02}$, ${\cal H}^{28}$ and ${\cal H}^{80}$ should be multiplied by $R$, whereas the remaining elements should be multiplied by $1-R$. Other elements can be obtained by complex conjugation. 

\end{widetext}
\end{appendix}

\end{document}